\documentclass[12pt]{article}
\makeatletter
\usepackage[margin=72.27pt]{geometry}
\parskip\z@
\usepackage[factor=700,stretch=14,shrink=14,step=1]{microtype}
\usepackage{enumitem}
\setlist{topsep=\smallskipamount,itemsep=\smallskipamount,
  parsep=\z@,partopsep=\z@,beginpenalty=0,midpenalty=0,
  endpenalty=0}
\usepackage[size=footnotesize,bf]{caption}
\usepackage{amsmath}
\usepackage{amssymb}
\allowdisplaybreaks
\usepackage{graphicx}
\usepackage{tabularx}
\usepackage{booktabs}
\PassOptionsToPackage{colorlinks=true,
  allcolors=blue,
  implicit=false}{hyperref}
\usepackage{orcidlink}
\usepackage{multirow}
\newcolumntype{C}{>{\centering\arraybackslash}X}
\newcolumntype{M}{>{\centering\arraybackslash$\displaystyle}X<{$}}
\def\E{\mathop{\kern\z@\mathbb E}}
\def\P{\mathop{\kern\z@\mathbb P}}
\def\tanhinv{\mathop{\operator@font tanh}^{-1}}

\AtBeginDocument{\baselineskip=\the\baselineskip plus 0.2pt minus 0.2pt}
\usepackage{amsthm}
\theoremstyle{definition}

\theoremstyle{plain}

\protected\def\bigskip{\vskip\bigskipamount}
\protected\def\medskip{\vskip\medskipamount}

\newdimen\@saveparindent
\newcount\@hd
\@hd\@ne
\def\section#1{\par\penalty\z@\bigskip
  \centerline{\textbf{\the\@hd. #1}}
  \global\advance\@hd\@ne
  \nobreak\medskip\nobreak
  \@saveparindent\parindent
  \parindent\z@
  \everypar{\parindent\@saveparindent}}
\def\fignum{\the\numexpr\thefigure+1\relax}

\def\linelabel#1{\textit{#1}---\ignorespaces}

\usepackage[authordate,doi=false,footmarkoff,
  maxbibnames=4,minbibnames=1,uniquelist=false,
  uniquename=false]{biblatex-chicago}

\let\revsdnamedelim\@empty
\bibhang\z@

\let\blx@err@patch\@gobble

\def\eq#1{equation~(\ref{#1})}
\def\Eq#1{Equation~(\ref{#1})}

\let\th@label\label
\def\newth#1#2#3{%
  \expandafter\def\csname th@#1\endcsname{%
    \begin{#2}\th@label{#1}#3\end{#2}}}
\def\th#1{\csname th@#1\endcsname}

\newth{scale-by-k}{theorem}
  {Suppose $k>0$, and let $V$ be the value function for an agent with instantaneous utility $u$. For an agent with utility $ku$, the value function is $kV$, and the saving rule remains unchanged.}

\newth{positive-increasing-V}{theorem}
  {When $t<T$ and $x>0$, $V$ is positive, strictly decreasing in $t$, and strictly increasing in $x$.}

\newth{scale-diff-eq}{lemma}
  {Let $y$ satisfy the differential equation $y'=f(t,y)$, and suppose $y(T)=0$. If $z$ satisfies the differential equation $z'=\lambda f(T+\lambda(t-T),z)$ with $\lambda>0$ and also has initial condition $z(T)=0$, then $z(t)=y(T+\lambda(t-T))$.}

\newth{scale-V}{theorem}
  {Changing $\lambda$ horizontally dilates $V$ around $T$. When $\lambda$ increases, the dilation is a shrink, and when $\lambda$ decreases, the dilation is a stretch.}

\newth{concave-V-single}{theorem}
  {When the agent has a single resource, V is concave.}

\newth{differential-eq-compare}{lemma}
  {Let $y$ and $z$ satisfy differential equations $y'=f(t,y)$ and $z'=g(t,z)$ with $f$ and $g$ negative, and suppose $y(T)=z(T)=0$. If $f<g<0$, then for $t<T$, $y(t)>z(t)$.}

\newth{normalize-deriv-int}{lemma}
  {Let $f$ and $g$ be increasing functions with continuous second derivative defined on the interval $[a,b]$. If $f$ is more concave than $g$, than for any $x$,
\[
\int_x^b\frac{f(s)-f(x)}{f'(x)}\:ds<\int_x^b\frac{g(s)-g(x)}{g'(x)}\:ds.
\]}

\newth{phi-psi-compare-single}{theorem}
  {Consider an agent with a single resource. If $w$ is more concave than $u$, then $V(t)<W(t)$ and $\psi(t)<\phi(t)$.}

\newth{discrete-envelope}{lemma}
  {Let $D$ be a discrete set, and let $I$ be an interval. Suppose that $f$ is a function defined on $[0,1]\times I\times D$ that is continuously differentiable in its first two arguments. Let $\beta\mathpunct{\colon}[0,1]\times I\to D$ be well behaved in the following sense: for any number $p$ in the interval $I$, the restriction $\beta|_p$ is piecewise constant on $[0,1]$, and for any point $(s,p)$ such that $\beta|_p$ is locally constant at $s$, $\beta$ is locally constant at $(s,p)$. Then
\[
\frac d{dp}\int_0^1 f(s,p,\beta(s,p))\:ds=
  \int_0^1\frac{\partial f}{\partial p}\,\bigg|_{s,p,\beta(s,p)}\:ds.
\]}

\newth{concave-max}{lemma}
  {Let $f$ and $g$ be strictly increasing and strictly concave functions defined on the natural numbers with $f(0)=g(0)=0$. Then $h(x)=\max_y\{f(x-y)+g(y)\}$ is also strictly increasing and, assuming generic $f$ and $g$, strictly concave. Further, $\max\{\Delta f(x),\Delta g(x)\}\leq\max\{\Delta f(x-y^*),\Delta g(y^*)\}=\Delta h(x)$, where $y^*$ is the optimum for $h(x)$. If $y^*$ satisfies $0<y^*<x$, then the inequality is strict.}

\newth{int-functions}{lemma}
  {If $\{f_i\}$ is a family of positive, increasing, discrete, and strictly concave functions indexed by $i\in[0,1]$, then
\[
\int_0^1f_i\:di
\]
is also positive, increasing, discrete, and concave.}

\newth{V-properties-divisible}{theorem}
  {Assume the agent's resources are divisible as described in this section. Then $V$ is concave in both arguments, and $\Delta V_t<0$. Here $V_t$ is the partial derivative of $V$ with respect to $t$, and $\Delta$ means discrete derivative with respect to $x$.}

\newth{y-properties}{corollary}
  {The saving rule $y$ is non-strictly decreasing with respect to $t$. (Equivalently, $x-y$ is increasing in $t$.) Both $y$ and $x-y$ are non-strictly increasing with respect to $x$.}

\newth{asymptote}{theorem}
  {For an agent with a divisible resource, $\lim_{t\to-\infty}V(t,x)=x\mu(1)$.}

\newth{preference-reversal}{theorem}
  {An agent with a divisible resource exhibits static preference reversal, and the time when the preference reversal happens is unique.}

\newth{correlation-aversion}{theorem}
  {With two random payments, the agent's expected value is strictly decreasing in the amount of correlation between the payments.}

\newth{monotonicity}{theorem}
  {At time $t$, the valuation of the agent's current resource stock plus a future payment at $\bar t$ strictly decreases as $\bar t$ gets larger starting from $\bar t=t$.}

\newth{log-concave-property}{lemma}
  {Let $f$ be a smooth and strictly increasing function with $f(0)=0$. If $f$ is log-concave, then for any $a$ such that $0<a<1$, $af'(f^{-1}(x))<f'(f^{-1}(ax))$.}

\newth{cutoffs-approaching}{theorem}
  {If $\zeta$ is log-concave, then $\phi_{i+1,j}-\phi_{i,j}$ is strictly decreasing for all times after $\phi_{i+1,j}^{-1}(1)$.}

\newth{more-concave-w-marginal}{theorem}
  {Consider an agent with a divisible resource. If $w$ is more concave than $u$ with respect to $\theta$, then for any $i$, $V_{i+1}-V_i<W_{i+1}-W_i$, where $W$ is the valuation that corresponds to $w$.}

\begin{document}

{\LARGE
\centerline{Reinterpreting Delay and Procrastination}
\centerline{Conrad Kosowsky*\kern0.1em\relax\orcidlink{0009-0007-0479-6746}}
\let\thefootnote\relax\footnotetext{*University of Michigan, Department of Economics and Center for the Study of Complex Systems. Email: coko@umich.edu.}
}

\vskip 0.3in

\begin{abstract}
\hyphenpenalty=500
\noindent I model a rational agent who experiences endogenous deadline pressure in the face of a fixed future deadline. The agent holds a resource stock, and opportunities to spend resources arise randomly according to a Poisson process. When the deadline is far away, the agent smooths consumption, but as the deadline approaches, the agent prioritizes current spending because of uncertainty about the future. The combination of concave utility and the agent's liquidity induces correlation aversion. Connecting the agent's risk and time preference is intuitive and leads to a model of procrastination where the agent overestimates their desire to spend resources.

\medskip

\end{abstract}

\section{Introduction}

Deadlines are ubiquitous. Projects have completion timelines; budgets expire; food goes bad. On a personal level, readers of this paper will be familiar with deadline pressure in any number of situations. In this paper, I investigate how deadlines affect incentives in the presence of randomness, and I model deadline pressure and procrastination. I consider a single rational agent who possesses a resource stock and moves forward through time until some fixed, known deadline $T$. The agent can consume resources only at specific spending opportunities, and the timing and quality of each opportunity are random and exogenous. For example, consider a funding agency that solicits proposals and distributes awards on a rolling basis until some internal deadline. The agency's budget is a resource stock, and each proposal is a spending opportunity. Whenever the agency receives a new proposal, it decides how much of its budget to allocate toward the proposal based on how well it aligns with the agency's aims. From the agency's perspective, proposals come in at random, and the funding agency does not know exactly how many it will receive before the deadline. Both the agent from my model and the funding agency of our example need to account for this uncertainty in spending opportunities if they want to consume resources optimally. This situation induces deadline pressure because the rational agent and the funding agency are compensating for the possibility that a given spending opportunity may be the last one.

In mathematical symbols, the agent's resource stock is $x$, and at any given time $t$, they encounter an opportunity to spend resources with probability $\lambda dt$. Formally, spending opportunities follow a Poisson process with rate parameter $\lambda$ on the interval $(-\infty,T]$, and after time $T$, the agent cannot spend more resources. In our example, $x$ is the size of the agency's budget, and $\lambda$ controls how quickly proposals arrive. If the agency is well-known and has an easy application process, we expect $\lambda$ to be relatively large, and if not, $\lambda$ may be smaller. The funding agency's budget expires at the deadline, so $x$ exogenously becomes 0 at $T$ in that case. If the deadline is public rather than internal, people will not submit proposals after $T$, and $\lambda$ also exogenously becomes 0. What's important here is that resources are useful only to the extent that the agent can spend them, so the valuation for the resource stock comes from possible future opportunities and is null from $T$ onward. In applications, we should expect at least one of $x$ or $\lambda$ to become 0 at $T$, but for the math in this paper, we remain agnostic about how the deadline manifests in regard to the agent's decision problem.

Whenever the agent reaches a spending opportunity, they first learn the quality $\theta$ associated with the opportunity, which is uniform on $[0,1]$ and independent of other opportunities. When our hypothetical funding agency receives a proposal, it reads the proposal to determine how the proposal matches the agency's aims, and higher $\theta$ means the proposal aligns with the agency's goals more closely. After learning $\theta$, the agent chooses $y$ resources to save, spends the remaining $x-y$ amount, and derives utility equal to $u(\theta,x-y)$, where $u$ is the agent's instantaneous utility function. The agent chooses $y$ to maximize the sum of current instantaneous utility and expected future utility from any remaining opportunities. Moving forward in time, the agent faces the same problem except with $y$ resources instead of $x$, so the agent's valuation of their resource stock solves a dynamic programming problem with $x$ and $t$ as state variables. In our example, this means that once the funding agency determines $\theta$ for a given proposal, it decides how much money to spend based only on $\theta$, its current budget, and the amount of time until $T$. Previous proposals do not factor into the decision, and such an approach is possible because of rolling monetary awards. If the agency released all of its funding decisions at the same time, its decision problem would look fundamentally different.

The assumptions of the previous paragraphs lead to a solution with rich properties that can inform our understanding of deadlines and delay. A solution to the model is a saving rule $y(\theta, t,x)$ that means the following: if the agent has $x$ resources at time $t$, and a spending opportunity arises with quality $\theta$, the agent saves $y(\theta, t,x)$ resources and spends $x-y(\theta, t,x)$. I abbreviate the saving rule to $y$ when the arguments are clear from context. Dual to $y$ is the agent's value function $V(t,x)$, which captures the expected sum of instantaneous utility from all future spending opportunities if the agent has $x$ resources at time $t$. Finding $y$ is equivalent to finding $V$, but as is typically the case in dynamic programming problems, $V$ is more tractable and will occupy more of our focus in this paper. One intuitive property of $y$ is that with all else being equal, the agent spends more at the next opportunity when the opportunity occurs closer to the deadline. In our example, if the funding agency receives the same proposal at different points in time and if its remaining budget is the same in both situations, it should spend more money on the proposal when it arrives later.

The Poisson process is central to the model solution and becomes particularly important near $T$. When the deadline is far away, the agent has a good sense of how many opportunities they will see between $t$ and $T$, and they essentially try to smooth consumption. Behavior in these situations resembles standard models of consumption smoothing from the macroeconomic literature, but as the deadline approaches, the agent departs from a consumption smoothing regime and prioritizes present consumption. The agent's uncertainty about the presence of a next opportunity drives their focus on current spending because they are compensating for the possibility that their current opportunity is the last one. In contrast to standard models of consumption smoothing, this effect becomes more pronounced as the agent reaches the deadline, and I consider it an example of endogenous deadline pressure.

A natural connection emerges between the optimal saving rule and the concavity of $u$. As is often the case in problems like this one, the Bellman equation commutes with multiplication by a constant (is homogeneous of degree 1), so scaling $u$ by a constant $k$ also scales $V$ by $k$ but has no effect on $y$. The agent's spending and saving choices are independent of the magnitude of $u$, so the concavity of $u$ controls $y$. The mathematical relationship between risk aversion and intertemporal elasticity of substitution is well known, and the decision theory literature regards it as puzzling. In my model, the relationship between concavity and $y$ is intuitive if we view it through the lens of ``waiting for the right moment.'' If $u$ is convex, the opportunity cost of spending with low $\theta$ is greater than if $u$ is more concave, so the agent will try to restrict their spending to opportunities with high $\theta$ values.

We can construct a model of procrastination using an agent who believes themself to be more pragmatic than they are. Suppose that instead of learning $\theta$ at an opportunity, the agent learns possible outputs of $u(\theta, x-y)$ for different values of $y$, so utility comes from the agent's experience of consumption rather than a predetermined ordering on a set of outcomes. If the agent believes they have instantaneous utility $\tilde u$ that is more concave with respect to $\theta$ than $u$, they infer a value $\smash{\tilde\theta}$ for the quality of the spending opportunity that is lower than $\theta$. The agent believes that they are encountering an excess of bad opportunities through random chance, and because they also use a decision rule based on $\tilde u$, the agent delays consumption more than they would if they knew $u$ and $\theta$ accurately. If our hypothetical funding agency claims to have a wide variety of interests but actually cares about only a very narrow set of proposals, it will mistakenly ascribe low $\theta$ values to many proposals and procrastinate on spending as a result.

The rest of this paper is organized as follows. Section~2 discusses related literature, and section~3 introduces the model in detail. Section~4 explains why the agent faces deadline pressure, and section~5 discusses the solution for arbitrary discrete resource stocks where all the pieces are the same size. Section~6 establishes that the agent exhibits correlation aversion and monotonicity with respect to payment timing. Section~7 connects risk and time preference including a model of procrastination, and section~8 concludes. This paper has two appendixes containing the derivation of the Bellman equation and proofs of all lemmas and theorems.

\section{Background}

This paper contributes to a variety of areas in decision theory and operations. My most significant contribution is to the literature on deadlines, but I also contribute to the literature on recursive preferences and correlation aversion, risk and time, and procrastination. This paper relates to work on the secretary problem, financial mathematics, and homogeneity of time preferences.

\linelabel{Deadlines} I model deadlines in broad terms. I formalize deadline pressure and show explicitly how it arises from time and opportunity constraints, and a rational agent in my model will bow to deadline pressure by discounting the future more heavily near $T$. The agent becomes less selective about opportunity quality as they move forward in time because the agent tries to unload their resource stock before running out of opportunities at the deadline. For fixed $t$, higher $\theta$ means higher spending, and the cutoff values of $\theta$ that separate different outputs for $y$ are decreasing in time. If the resource stock is discrete, then sufficiently close to $T$, the agent spends their entire resource stock whenever an opportunity arises with $\theta$ greater than some threshold. The threshold is decreasing in time and equals 0 at $T$.

Models involving a deadline and a stochastic cost or benefit for a task are already present in the literature. \textcite{heidhues-strack-2021} and \textcite{hyndman-bisin-2022} both consider a single rational agent who solves an optimal stopping problem to decide when to complete a task, and either the cost or reward from the task varies randomly. \textcite{heidhues-strack-2021} consider a quasi-hyperbolic agent who may or may not be naive, and \textcite{hyndman-bisin-2022} consider agents who exhibit both exponential and sophisticated or naive hyperbolic discounting. If $x$ is a single, indivisible resource, my model becomes intuitively very similar to the model of either paper except with a different discounting scheme for the agent. My work differs from both papers in my aims. I analyze the incentives created by deadlines, whereas \textcite{heidhues-strack-2021} focus on identifiability of the agent's discount parameters, and \textcite{hyndman-bisin-2022} investigate demand for commitment devices in the form of deadlines.

Deadlines usually appear in the context of some other topic. This literature is too broad to survey in its entirety, but the following papers form a slice of the literature. Of particular interest is the effect of deadlines on job performance. \textcite{lambert-et-al-2017} examine the effects of deadlines on audits and conclude that more time pressure results in lower audit quality. See \textcite{kuutila-et-al-2020} for a review of the literature on deadlines, time pressure, and software engineering. Deadlines come up in the theoretical literature, particularly in mechanism design \autocite{arefeva-meng-2017, green-taylor-2016, madsen-2022, mierendorff-2016}. \textcite{karagozoglu-kocher-2019} examine how time pressure affects bargaining in an experiment. \textcite{bizzotto-rudiger-vigier-2021} construct an information design problem with a deadline and exogenous information. \textcite{coey-larsen-platt-2020} model a consumer-search problem with a deadline and find that as the deadline approaches, the consumer will pay more for a good. The intuition in all of these papers is similar to my findings, where the agent will spend more resources on lower-quality opportunities when under time pressure. See also \textcite{emanuel-katzir-liberman-2022} for discussion of this idea from a psychological point of view.

\linelabel{Liquidity} The agent in this paper exhibits monotonicity with respect to payment timing. \textcite{blavatskyy-2016} points out that in standard intertemporal choice models, an agent may prefer to delay a portion of a payment scheduled for today. Suppose the agent discounts a future payment by $\delta$. (With only two payments, we can specify $\delta$ while being agnostic about the form of the agent's discounting.) If the agent receives \$1 today and \$1 at a later time, the utility from these payments is $u(1)+\delta u(1)$. If $u$ is concave, and $\delta$ is large enough, the sum is greater than $u(2)$, even though it makes no sense for the agent to delay a monetary payment without reason. Monotonicity with respect to payment timing means that moving all or part of a payment later never helps the agent. \textcite{blavatskyy-2016} induces monotonicity by modeling intertemporal choice with rank-based utility, and I induce monotonicity by allowing the agent to hold resources resources until the next spending opportunity.

I am implicitly providing the agent with some liquidity, and this assumption is significant for understanding and testing how people prioritize the timing of payments and consumption. In a review of experiments where subjects decide whether to receive money sooner or later, \textcite[][p.\ 300]{cohen-et-al-2020} write that such experiments are ``assuming that a monetary payment at date $t$ generates a dollar-equivalent incremental consumption event at time $t$\dots\ This assumption is inconsistent both with the empirical evidence and with standard economic theory. Only perfectly liquidity-constrained consumers or perfectly myopic consumers would instantly consume every payment they receive.'' This paper contrasts with that experimental assumption because payments to the agent provide value through future spending opportunities rather than by going immediately into the agent's instantaneous utility function.

\linelabel{Correlation Aversion} Second, I construct a rational agent whose valuation of resources is additive with respect to future instantaneous utility yet who nevertheless exhibits correlation aversion. Correlation aversion means that if an agent receives multiple random payments, the agent prefers their outcomes be less correlated, and this behavior is both desirable for a rational agent and typically associated with non-additive value functions. Two of the most common preference models in the intertemporal choice literature are additive discounting and Epstein-Zin preferences.%
%
%
\footnote{Additive discounting dates back to \textcite{samuelson-1937}. \textcite{kreps-porteus-1978} introduced a framework for modeling preference for timing of the resolution of uncertainty, and \textcite{epstein-zin-1989} developed these ideas into a recursive model of individual preference. Additive discounting has received scrutiny in part because it cannot distinguish correlations between payment probabilities at different time periods. See \textcite{aase-2021} and \textcite{stanca-2023} for more discussion of this point.

It may be helpful for some readers to see a definition of both types of preference models. Let $c=(c_1,c_2,\dots)$ be a sequence of nonnegative random variables, where each sequence entry represents consumption in a different time period, and suppose $c_0$ is current, known consumption. Our task is to assign a real number $U$ to $(c_0,c_1,c_2,\dots)$ that encodes the value of the consumption stream. If an agent exhibits additive discounting over a discrete and infinite time-horizon, that means we can express their valuation as
\[
U(c_0,c)=u(c_0)+\E\left[\sum_{t=1}^\infty D(t)u(c_t)\right],
\]
\noindent where $D$ is their discount function and $u$ is their instantaneous utility function. In particular, exponential discounting means $D(t)=\delta^t$, and hyperbolic discounting means $D(t)=(1+\delta t)^{-1}$.

Notice that $U$ from the previous paragraph maps a real number and sequence of random variables to a real number, so $U(c)$ is a random variable. This is because $f(x)=U(x,c_2,c_3,\dots)$ maps real numbers to real numbers, and therefore $U(c)=f(c_1)$ is a transformation of a random variable. Let $\rho$ and $\sigma$ be increasing diffeomorphisms from the nonnegative reals to the nonnegative reals. If an agent exhibits Epstein-Zin preferences over a discrete and infinite time-horizon, that means their valuation satisfies the recursion relation
\[
U(c_0,c)=\rho^{-1}\left(\rho\circ u(c_0)+
  \delta\rho\circ\sigma^{-1}\E\left[\sigma\circ U(c)\right]\right),
\]
where $\delta$ is a discount parameter. Traditionally, $\rho$ and $\sigma$ are power functions, and in \textcite{epstein-zin-1989}, $u$ is the identity function. Epstein-Zin preferences allow us to better control how an agent aggregates utility over multiple time periods.}
Additive discounting is simple, has been standard in the literature for many decades, and cannot explain correlation aversion, whereas Epstein-Zin preferences are newer, more sophisticated, and do support correlation aversion. In terms of simplicity, the agent of this paper has a value function that lies somewhere between additive discounting and Epstein-Zin preferences, but the agent still exhibits correlation aversion.\footnote{As is the case with Epstein-Zin preferences, the agent in my model has a value function $V$ that is recursive. However, the recursive preferences of this paper are not Epstein-Zin.} It follows that a non-additive valuation is not the simplest model that explains this phenomenon.

Recent papers on recursive preferences encompass a variety of topics in decision theory, usually relating to preference for information or resolution of uncertainty. Whenever I say recursive preferences in this paper, I mean any value function that arises from a dynamic programming problem. Epstein-Zin is the most well-known form of recursive preferences, but it is not the only formulation. \textcite{al-najjar-shmaya-2019} consider a rational agent with Epstein-Zin preferences whose discount factor approaches 1, and they show in closed form that the agent exhibits ambiguity aversion. \textcite{evren-2019} investigates ambiguity aversion in agent with recursive but not Epstein-Zin preferences. \textcite{li-2020} models agents who avoid partial information. \textcite{falk-zimmermann-2016} run an experiment where subjects usually prefer information sooner depending on context, and \textcite{brown-guo-je-2022} and \textcite{nielsen-2020} conduct experiments to test demand for non-instrumental information. In a related theoretical contribution, \textcite{gul-natenzon-pesendorfer-2021} introduce the concept of random lotteries to model agents who prefer non-instrumental information. \textcite{bommier-kochov-le-grand-2017} classify recursive preference models that are monotone in payment possiblities, and \textcite{kochov-song-2023} analyze infinitely repeated games when the agents have recursive preferences.

\linelabel{Risk and Time} My model provides an intuitive connection between risk and time preference because we can interpret a more risk-seeking attitide with respect to $\theta$ as ``waiting for the right moment,'' a motivation that will undoubtedly be familiar to many readers. A more risk-seeking agent spends resources sooner when they see high values of $\theta$ and waits if $\theta$ is lower. The relationship between risk and time has attracted attention for decades, and a big motivation behind Epstein-Zin preferences is their ability to separate an agent's attitudes toward risk and time. The literature is currently split on whether risk and time preference ought to affect each other. For example, \textcite[][p.\ 2]{de-castro-et-al-2023} call it ``a drawback'' that ``in the [expected utility] model, risk aversion and [intertemporal substitution] cannot be disentangled,'' but \textcite[][p.\ 311]{epper-fehr-duda-2024} claim that ``[a] considerable body of experimental evidence suggests\dots\ that risk taking and time discounting are linked.'' By providing a straightforward explanation for why risk and time preference affect each other here, I suggest that it is plausible to connect risk and time more generally. A similar flavor of procrastination arises when the agent misperceives the passage of time.

A swath of recent papers have attempted to experimentally determine subjects' coefficient of risk aversion and elasticity of intertemporal substitution. \textcite{andersen-et-al-2008} first pointed out the importance of measuring risk and time attitudes simultaneously, and \textcite{andreoni-sprenger-2012} built on this idea when they introduced convex time budgets, an experimental technique that has become popular throughout the literature. A large number of authors have adopted or expanded the approaches in these two papers \autocite{andersen-et-al-2018,apesteguia-ballester-gutierrez-2019,balakrishnan-haushofer-jakiela-2020,bernedo-del-carpio-alpizar-ferraro-2022,blavatskyy-maafi-2020,cheung-2020,de-castro-et-al-2023,sornasundaram-eli-2022,sun-potters-2022}. Some recent work has attempted to better fit data through prospect theory and probability weighting \autocite{diecidue-hardardottir-islam-2023,holden-tilahun-sommervoll-2022,kemel-paraschiv-2023,lampe-weber-2021}. Finally, a few authors have specifically attempted to measure risk and time preference without relying on a preference model by measuring timing premia in the case of \textcite{meissner-pfeiffer-2022} or correlations in the case of \textcite{rohde-yu-2023}.

Theoretical work in this area has focused on the relationship between risk and time under different models. \textcite{fahrenwaldt-jensen-steffensen-2020} and \textcite{lau-2019} formulate methods to separate risk and time preference using certainty equivalents. \textcite{chakraborty-halevy-saito-2020} and \textcite{saito-2015} use a general framework to make statements about behavior of an agent who believes that the future is inherently uncertain. \textcite{epper-fehr-duda-2024} model an agent using rank-dependent utility in an attempt to explain seven different observations about human behavior. See \textcite{ericson-laibson-2019} and \textcite{thimme-2017} for reviews of the literature on intertemporal choice.

\linelabel{Procrastination} If the agent misperceives their instantaneous utility function, we end up with a model of procrastination. I imagine that an agent overestimates their eagerness to complete a task and is ``waiting for the right moment'' more than they believe themself to be. This idea differs from traditional explanations of procrastination in economics, where present bias arises from a time-inconsistent utility function such as hyperbolic or quasi-hyperbolic discounting. For example, \textcite{augenblick-rabin-2019}, \textcite{bisin-hyndman-2020}, and \textcite{fedyk-2024} measure present bias in experiments where they model subjects using quasi-hyperbolic discounting. \textcite{niu-2023} models an agent with quasi-hyperbolic discounting who allocates a continuous flow of effort to a task between a starting time and a final deadline. In these types of papers, agents experience tension between what they plan and what they do, and commitment devices become an important tool for understanding how agents behave. In my model, a procrastinating agent has no interest in a commitment device.

This paper contributes to the more recent literature that explores agent beliefs as a motive for present bias and procrastination. \textcite{brunnermeier-et-al-2017} model an optimistic agent whose beliefs lead to delay in behavior, and \textcite{breig-gibson-shrader-2020} and \textcite{cordes-friedrichsen-shudy-2024} run experiments where they conclude that optimistic subjects delay task completion. \textcite{kaufmann-2022} and \textcite{zhang-2021} study projection bias and how projecting beliefs or tastes into the future affects an agent's time preference. \textcite{heidhues-koszegi-strack-2024} analyze an agent who misperceives their objective function in past time periods, which is similar to how I model procrastination. The agent views their pervious motives as more benevolent than they actually were and ends up exhibiting present bias. The authors allow the agent to learn about their misperceived objective function, whereas in this paper, an agent who procrastinates cannot gain self-awareness so easily. In the psychology literature, \textcite{jagga-srinivasan-srivastava-2021} and \textcite{zhang-feng-2020} conduct experiments where subjects overrate their future interest in a task and procrastinate as a result.

\linelabel{Secretary Problem} Apart from the decision theory literature, my model has a similar flavor to the secretary problem from optimal stopping theory. Suppose an academic department wants to hire a new staff member. The faculty member in charge of hiring interviews the candidates sequentially, and they learn whether or not the current candidate is a better fit than all previous applicants. The department must make an offer immediately after the interview, or the candidate finds a job elsewhere. In the traditional secretary problem, the department wants to maximize its chance of hiring the best candidate, and when the number of applicants gets large, the optimal strategy is to interview the first $1/e$ candidates and accept the next applicant who is better than all previous applicants. My model differs from the secretary problem in that the agent of this paper knows the distribution of $\theta$, spends multiple resources at a time, and maximizes expected utility instead of finding the best spending opportunity before $T$.

The literature on this topic is abundant, and I cover only a small portion of it here. \textcite{bearden-2006} solves a variant of the secretary problem where the hiring committee maximizes utility, which is one of the most conceptually similar papers to mine from the literature. Recent work on the secretary problem has focused on extending the results to different objectives or orderings of the candidates. \textcite{albers-ladewig-2021} provide new results for the problem of selecting multiple candidates. \textcite{hoefer-kodric-2017} explore the secretary problem in more general contexts such as picking subsets of a graph or a matroid. \textcite{jones-2020}, \textcite{liu-milenkovic-2022}, and \textcite{liu-milenkovic-moustakides-2023} consider variants of the problem where the order of candidates follows a Mallows distribution. \textcite{gnedin-2022} suggests improvements to the traditional $1/e$ solution, and \textcite{demers-2019} investigates the stopping time of the optimal solution.

\linelabel{Financial Mathematics} Some work in financial math approaches timing of events similarly to this paper. \textcite{bayraktar-ludkovski-2010} model inventory management where the timing of orders for products follows a compound Poisson process with rate parameters determined by a Markov process. \textcite{bayraktar-ludkovski-2011} and \textcite{bayraktar-ludkovski-2011} derive optimal strategies for trading in financial markets where order flow follows a Poisson process, with specific focus on liquidation of large securities. For a review of this type of literature, see \textcite{donnelly-2022}.

\linelabel{Homogeneity of Time Preference} It is also worth being clear about the relationship between time and time discounting in my model. Time preferences can be homogeneous with respect to the passage of time in three ways: stationarity, time invariance, and time consistency. We can most easily understand these properties in terms of an evaluation period (the agent's present time) and a consumption period (a known future time for the agent). During the evaluation period, the agent chooses one of two alternatives to consume either upon reaching the consumption period or possibly at a fixed, known amount of time after the consumption period. Stationarity means the agent chooses independently of the consumption period, and time consistency means the agent chooses independently of the evaluation period. Time invariance means that if we change both the evaluation period and the consumption period by the same amount, the agent's choice remains the same. See \textcite{halevy-2015} for discussion of these properties. \textcite{koopmans-1960} introduced stationarity, and \textcite{halevy-2015} formalized time invariance. I am less clear on the origins of time consistency, but the concept dates back at least to \textcite{samuelson-1937}.

These properties affect the agent's behavior by prohibiting preference reversals. A preference reversal means that an agent makes two different choices depending on the timing of either the decision or the consumption. A static preference reversal is one that results from changing the consumption period (while fixing the evaluation period), and a dynamic preference reversal is the opposite. Dynamic preference reversals have received more attention than static preference reversals, to the point that some authors omit the word ``dynamic'' and use the general term ``preference reversal'' to refer to what I call dynamic preference reversal. An agent with stationary time preferences never exhibits static preference reversal, and an agent with time consistent time preferences never exhibits dynamic preference reversal. Time invariance makes both types of preference reversals equivalent. See \textcite{chen-et-al-2019} for more discussion of static versus dynamic preference reversals. \textcite{halevy-2015} showed that any two properties imply the third.

Because I study deadlines, my model by definition violates time invariance, and I show later in this paper that my model also violates stationarity. The agent's value function comes from a Bellman equation, so the agent exhibits time consistency. In choosing between a smaller-sooner or later-larger payment, the agent prefers the later-larger payment away from the deadline and the smaller-sooner payment near the deadline. Far away from the deadline, the agent recognizes that they will eventually switch to a smaller-sooner payment, so the preference reversal is static. The preference reversal is not dynamic because it happens in a time consistent fashion. Once the agent reaches the time period when they expect to switch to a smaller payment, they actually do so.

\section{The Model}

\begin{figure}[t]
\def\arraystretch{1.1}
\centerline{\bfseries Figure \fignum: Notation\strut}
\centerline{\begin{tabularx}{\textwidth}{lX<{\raggedright\arraybackslash}
  lX<{\raggedright\arraybackslash}}\toprule
Notation & Meaning & Notation & Meaning\\\midrule
$u$, $w$ & Instantaneous utility function & $\zeta$, $\xi$ & $\theta$-component of $u$ or $w$ \\
$V$, $W$ & Value function & \smash{$\tilde\zeta$} & Misperceived $\zeta$ \\
$y$, $y^*$ & Saving rule (model solution) & $\mu$ & $x$-component of $u$ or $w$ \\
$t$ & Current time & $x$ & Resource stock \\
$\bar t$ & Future time & $\bar x$ & Future payment\\
$T$ & Deadline & $\theta$ & Opportunity quality\\
$\lambda$ & Poisson rate & \smash{$\tilde\theta$} & Misperceived $\theta$ \\
\smash{$\tilde\lambda$} & Misperceived $\lambda$ & $c$ & Parameter controlling correlation of random payments\\
\smash{$\tilde V$} & Value function with multiple payments or misperceived $V$
  & $\phi_{i,j}$ & The $j$th cutoff for an agent with $i$ resources \\\bottomrule
\end{tabularx}}
\caption{Summary of notation used in this paper}
\end{figure}

In this section, I define the model precisely and establish some straightforward properties of the solution. See figure~1 for a summary of notation used in this paper. We consider a rational agent with a resource stock $x$. The agent spends their resources at various points in time and derives utility according to their instantaneous utility function $u$. By definition, $u$ takes two arguments, one for the quality $\theta$ of the spending opportunity and one for the amount $x$ of resources spent. We assume that $u(0,x)=u(\theta,0)=0$ and that $u$ is smooth and strictly increasing in $x$ and $\theta$. We also assume that $u$ is multiplicatively separable, so we can write $u(\theta,x)=\zeta(\theta)\mu(x)$. This technical assumption means that adjusting the quantity of consumption changes how opportunity quality affects utility in a way that is the same at all quality levels and vice versa. Restricting attention to separable utility functions comes with some loss of generality, but I claim the model is still very general. We have an extra degree of freedom because $\zeta$ and $\mu$ can scale inversely to one another, so we assume without loss of generality that $\zeta(1)=1$. Finally, we assume that $\mu$ is concave in $x$. We do not make any assumptions about concavity of $\zeta$, but theorem~\ref{cutoffs-approaching} applies only when $\zeta$ is log-concave.

Time itself is continuous, but opportunities to spend the resources are discrete and follow a Poisson process with rate parameter $\lambda$. The Poisson allows the resource stock to jump discretely in a continuous-time environment, which is a conceptual advantage relative to many models in the literature. Whenever the agent reaches a new opportunity, they learn the quality parameter $\theta$ associated with the opportunity and then decide what to save and what to spend. We assume that $V(t,x)$ encodes the sum of expected instantaneous utility from all future spending opportunities, so if the agent spends $x-y$ and saves $y$ resources at the current opportunity, they derive $u(\theta,x-y)$ utility immediately and $V(t,y)$ value from future opportunities. It follows that the expected value of an opportunity at (future) time $t$ is
\begin{equation}
\E\max\{u(\theta,x-y)+V(t,y)\},
\label{e-max}
\end{equation}
where we maximize over possible choices of $y$ and then take expectation with respect to $\theta$. The maximum operator occurs inside the expected value operator because the agent learns $\theta$ before choosing $y$.

Because the agent faces the same decision problem at each time $t$, we can restyle the agent's choice as a dynamic programming problem. To write the Bellman equation, we note that the current value of the agent's resources must equal the expected sum of instantaneous utility from the next spending opportunity plus the value of whatever remains afterward. \Eq{e-max} is exactly this quantity if the next spending opportunity occurs at some fixed time $t$, but in general, the timing of the next opportunity is random and could happen at any point before $T$. It follows that the agent's valuation is the expected expected value of \eq{e-max} with respect to $t$, where we weight different times according to their probability of being the next opportunity. For a Poisson process with rate parameter $\lambda$, the waiting time until the next event is an exponential distribution with rate parameter $\lambda$, so the Bellman equation for $V$ is
\begin{equation}
\label{bellman}
V(t,x)=\int_t^T\E\max_y\{u(\theta,x-y)+V(s,y)\}\lambda e^{\lambda(t-s)}\:ds.
\end{equation}
See Appendix~A for a more formal description of $V$. \Eq{bellman} completely determines the agent's behavior. The upper limit of the integral is $T$ because the agent cares only about opportunities before $T$, so $V(T,x)=0$. It follows that $V(t,0)=0$. Because expected value and maximum operators commute with multiplication by a positive constant, we have the following theorem.

\th{scale-by-k}

Differentiating \eq{bellman} gives us a differential equation for $V$. Applying the fundamental theorem of calculus, we get
\begin{equation}
\frac{dV}{dt}\,\bigg|_{(t,x)}=\lambda\left[V(t,x)-\E\max_y\{u(\theta,x-y)+V(t,y)\}\right],
\label{differential-eq}
\end{equation}
and because $V(T,x)=0$, it follows that $V_t(T,x)=-\lambda\E u(\theta,x)<0$, where $V_t$ is the partial derivative of $V$ with respect to $t$. This means there exists a neighborhood of $T$ where $V$ is positive, and we can extend the result to all $t<T$ by applying continuous induction to \eq{bellman}. By considering the maximand of \eq{differential-eq} with $y=x$, we see immediately that $dV/dt\leq0$. Comparing different values of $y$ allows us to make this inequality strict and establish monotonicity with respect to $x$. We have the following theorem.

\th{positive-increasing-V}

The Poisson rate parameter determines the frequency of spending opportunities, so we can think of $\lambda$ as controlling the agent's perception of time. When $\lambda$ is large, the agent sees many spending opportunities, and $V$ changes relatively little over a given time interval. When $\lambda$ is small, the agent sees few spending opportunities, and $V$ decreases much more. Mathematically, we see from \eq{differential-eq} that changing lambda scales the differential equation, and we have the following results.

\th{scale-diff-eq}

\th{scale-V}

Up to this point in the paper, I have avoided distinguishing between continuous and discrete resource stocks because the intuition behind the model is the same in both cases. In the next section, I treat $x$ as a continuous stock because the Euler equation makes the effect of the Poisson process transparent, and in the remaining sections of this paper, I assume $x$ is discrete because many obvious resource stocks, such as money, are discrete. We assume that $x$ is divisible into $n$ pieces with cutpoints $0=x_0<x_1<\dots<x_n=x$, and to keep things simple, we assume $x_i\propto i$. This means all pieces of $x$ are the same size, so it doesn't matter what order the agent spends the pieces in, only how many they spend at a time.\footnote{If the pieces of $x$ are different sizes, the problem very quickly becomes NP-hard. That is a Bad Thing. See also \textcite{babaioff-et-al-2007} and \textcite{babaioff-et-al-2009} for discussion of the secretary problem where candidates have weights. See also \textcite{cacchiani-et-al-2022a} and \textcite{cacchiani-et-al-2022b} for discussion of the knapsack problem more generally.} 
The scale of $x$ doesn't matter, so we assume without loss of generality that $x_i=i$. We make use of discrete derivatives with respect to $x$, and we let $\Delta$ be the discrete derivative operator. For a function $f$, $\Delta f(x)=f(x+1)-f(x)$, and when I use $\Delta$ in the rest of this paper, it will always mean discrete derivative.

\section{Deadline Pressure}

The Poisson process is the driving factor that creates a sense of deadline pressure for the agent, and we can most easily see how this happens by examining the Euler equation. The Euler equation only makes sense for continuous $x$, so in this section, in contrast to the rest of the paper, we treat $x$ as a continuous variable. In my model, opportunities follow a Poisson process, and \eq{bellman} is the Bellman equation. Consider a variation of my model without a Poisson process. Time is discrete, but opportunities still have some exogenous quality $\theta$. In this case, the Bellman equation simplifies to
\begin{equation}
V(t,x)=\E\max\{u(\theta,x-y)+V(t+1,y)\}.
\label{no-poisson-bellman}
\end{equation}
Differentiating equations~(\ref{bellman}) and (\ref{no-poisson-bellman}) with respect to $x$ and applying the envelope theorem gives us
\begin{align*}
V_x(t,x)&=\int_t^T\E u_x(\theta,x-y^*)\lambda e^{\lambda(t-s)}\:ds
  &\text{\clap{and}}&&
  V_x(t,x)&=\E u_x(\theta,x-y^*)
\end{align*}
for both versions of the model, where $y^*$ is the optimal value of $y$.

Consider the agent at time $t$ after they learn the value of $\theta$. In the case of no Poisson process, the first-order condition gives us
\[
u_x(\theta,x-y^*)=V_x(t,y^*)=\E u_x(\theta,y^*-z^+),
\]
where $y^*$ is the optimal amount to save at time $t$ and $z^+$ is the optimal amount to save at $t+1$. It follows that the Euler equation is
\begin{equation}
\frac{\E u_x(\theta,y^*-z^+)}{u_x(\theta,x-y^*)}=1.
\label{euler-no-poisson}
\end{equation}
The agent attempts to smooth consumption by equating marginal utilities in expectation. With exponential discounting using discount factor $\beta$, the right side becomes $1/\beta$. For the model with the Poisson process, the same calculation gives us
\[
u_x(\theta,x-y^*)=V(t,y^*)=
  \int_t^T\E u_x(\theta,y^*-z^+)\lambda e^{\lambda(t-s)}\:ds
  =\E\left[u_x(\theta,y^*-z^+);t^+\leq T\right],
\]
where $t^+$ is the time of the next opportunity and $z^+$ is the optimal amount to save at that time. Here the $\E$ operator in the integrand is an expectation with respect to $\theta$, whereas the $\E$ operator in the right-most term means expectation with respect to both $\theta$ and $t^+$. The $t^+\leq T$ restriction is mostly notational and emphasizes that we can never guarantee a next opportunity for the agent---with some positive probability, the current spending opportunity is the agent's last one.

The two Euler equations are similar, but we can condition on a next opportunity happening, and then things become very different. We have
\begin{align}
\frac{\E\left[u_x(\theta,y^*-z^+);t\leq T\right]}{u_x(\theta,x-y^*)}&=1\nonumber\\
\frac{\E\left[u_x(\theta,y^*-z^+)\middle|t\leq T\right]
  \P(t^+\leq T)}{u_x(\theta,x-y^*)}&=1\nonumber\\
\frac{\E\left[u_x(\theta,y^*-z^+)\middle|t\leq T\right]}{u_x(\theta,x-y^*)}
  &=\frac1{\P(t^+\leq T)}=\frac{1}{1-e^{\lambda(T-t)}}.
\label{euler}
\end{align}
When $T-t$ is large, the right side of \eq{euler} is close to 1, and the equation is very similar to \eq{euler-no-poisson}. However, conditional on a next opportunity happening, the ratio of the agent's marginal utilities becomes infinite close to $T$. The agent wants to spend the entire resource stock before $T$, and the incentive to do so gets larger and larger as $T$ approaches. This situation stands in contrast to the discrete-time model without the Poisson process. Even if we impose exponential discounting, which means both types of agents will prioritize current consumption, the discrete-time agent's desire to spend resources now does not change as the deadline approaches, and near $T$, the discrete-time agent never prioritizes current consumption as much as the agent with the Poisson process.

\Eq{euler} drives the agent's behavior in this model, and it means the agent exhibits endogenous, time-dependent discounting that becomes infinitely large near $T$. This behavior is due to the agent's uncertainty about the presence of a next opportunity, and we can see the effect more clearly by exogenously imposing a similar discounting regime in the discrete-time model. Let $f$ be a function that is always less than 1 and is positive for $t<T$. If $f$ encodes the discount factor for the next time period relative to the current time period, the Bellman equation for the discrete-time model becomes
\[
V(t,x)=\E\max\{u(\theta,x-y)+f(t+1)V(t+1,y)\}.
\]
The first-order condition is now $u_x(\theta,x-y^*)=f(t+1)V(t+1,y^*)$, and repeating the previous calculation gives us a new Euler equation of the form
\begin{equation}
\frac{\E u_x(\theta,y^*-z^+)}{u_x(\theta,x-y^*)}=\frac1{f(t+1)}.
\label{euler-discrete-discounting}
\end{equation}
It follows that the discrete-time model without Poisson process is analogous to the model with Poisson process if $f(t)=1-e^{\lambda(T-t)}$.

We should expect similar behavior from any agent whose discounting resembles the incentives from the Poisson process. For example, if $f(T)=0$, the right side of \eq{euler-discrete-discounting} becomes infinite, and the agent spends all remaining resources at time $T-1$. We can see this behavior clearly from the Bellman equation because
\begin{align}
V(T-1,x)&=\E\max\{u(\theta,x-y)+f(T)V(T,y)\}\label{bellman-t-1}\\
&=\E\max\{u(\theta,x-y)\},\nonumber
\end{align}
which has solution $y^*=0$. By contrast, in the discrete-time model without discounting or with constant exponential discounting, the agent is happy to save resources in the penultimate period because the Bellman equation for $V(T-1,x)$ no longer contains the $f(T)$ factor from \eq{bellman-t-1} that made $V(T,y^*)$ vanish. If $f(t)$ is close to 1 away from $T$, the agent smooths consumption exactly as in the Poisson process case. Because both the Poisson process and a specific type of discounting result in comparable Euler equations and like behavior, we can conceptualize the Poisson process as creating endogenous discounting that depends heavily on the deadline. That is exactly how deadline pressure works.

\section{Solution Properties}

In this section, we analyze the solution to \eq{bellman} in more detail. We assume that for some natural number $n$, $x=n$, and we can divide $x$ into $n$ unit-sized pieces. The Bellman equation is more amenable to investigation if we write it as a differential equation, and we will describe most of approach in terms of \eq{differential-eq}. In regard to finding actual solutions, \eq{differential-eq} becomes intractable very quickly, but we can solve the simplest cases in closed form. Consider an agent with a single, indivisible resource, and suppose that $\zeta(\theta)=\theta$. When $n=1$, the value of $\mu(1)$ doesn't matter as long as it is positive, so we assume for this example without loss of generality that $\mu(1)=1$. The differential equation for $V$ simplifies to
\[
V_t(t,1)=\lambda(V(t,1)-\max\{\theta, V(t,1)\}),
\]
where $V_t$ is the partial derivative of $V$ with respect to $t$.

On the probability space $[0,1]$, the function $V(t,1)$ is a constant, and the function $\theta$ is the identity. We have
\begin{align*}
V_t&=\lambda\left(V-\int_0^VV\:d\theta-\int_V^1\theta\:d\theta\right)\\
&=\lambda\left(V-V^2-\left(\frac12-\frac12V^2\right)\right)\\
&=\lambda\left(-\frac12+V-\frac12V^2\right)=-\frac12\lambda(1-V)^2.
\end{align*}
This differential equation is separable and has a closed-form solution given by
\[
V(t,1)=1+\frac{2}{\lambda t+C},
\]
where $C$ is a constant. Under the initial condition $V(T,1)=0$, we have $C=-2-\lambda T$, and the solution becomes
\[
V(t,1)=1-\frac{2}{\lambda(T-t)+2}.
\]
This function is positive and decreasing, as we expect, and it is concave, which will be true of $V$ in general. Figure~2 shows an example of this  Unfortunately, for more complicated $u$, \eq{differential-eq} very quickly becomes impossible to solve explicitly, although it is amenable to numerical methods.\footnote{If $u(\theta)=\theta^k$, we end up with
\[
V'=\lambda\left(V-\frac k{k+1}\sqrt[k]{V^{k+1}}-\frac1{k+1}\right)
\]
for $V$. When $k=2$, a computerized differential equation solver gives
\[
C-\frac{\lambda t}3=\frac2{3-3\sqrt{V}}-\frac49\tanhinv\left(\frac2{3\sqrt{V}}+\frac13\right)
\]
as the solution to the differential equation. It is impossible to invert this implicit definition for $V$.}

\begin{figure}[t]
\centerline{\bfseries Figure \fignum: Example Solution}
\smallskip
\centerline{\includegraphics{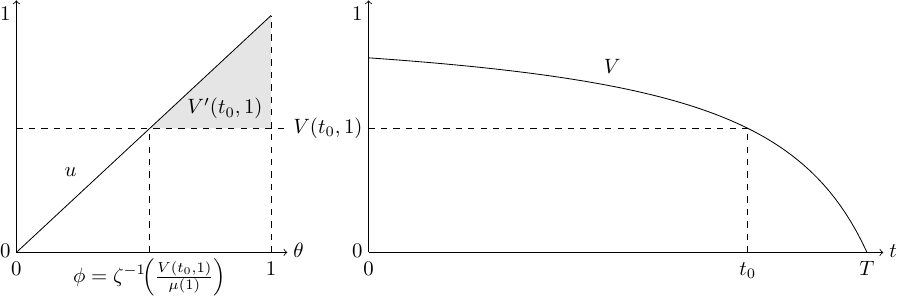}}
\caption{An example solution with $\zeta(\theta)=\theta$. The left graph shows $u(\theta,1)=\zeta(\theta)$, and the right graph shows $V(t,1)$. The shaded gray region corresponds to $V'(t_0,1)$, and the size of that region is the magnitude of $V'$. Notice that as $V(t,1)$ decreases, the size of the shaded region must increase, which explains intuitively why $V$ is concave.}
\end{figure}

It is instructive to consider the single-resource case with arbitrary $u$ before tackling the multiple-resource case. At a spending opportunity with quality $\theta$, the agent compares $V(t,1)$ and $u(\theta,1)$, and they spend their resource whenever $\theta$ exceeds a cutoff value $\phi$, where $\phi$ satisfies
\begin{align*}
u(\phi,1)=\zeta(\phi)\mu(1)&=V(t,1)\\
\phi&=\zeta^{-1}\left(\frac{V(t,1)}{\mu(1)}\right).
\end{align*}
Because $\theta$ can be between 0 and 1, the expected value in \eq{differential-eq} is exactly $V(t,1)$ if $V(t,1)=u(1,1)=\mu(1)$, so the differential equation for $V$ has a nullcline at $\mu(1)$. If $V<\mu(1)$, then for some set of $\theta$-values with positive measure, $V<u(\theta,1)$, so $V$ is decreasing in this case. It follows that the value function has an asymptote at $\mu(1)$ as $t$ gets big negative. The more time an agent has before the deadline, the more likely they are to find an opportunity with $\theta$ very close to 1. Further, we can interpret the time derivative of $V$ as negative the portion of the undergraph of $u(\theta,1)$ that lies above $V$ and between $\psi$ and 1. As $t$ increases, the area of this region enlarges, and we have the following result.

\th{concave-V-single}

The single-resource case is simple enough that we can connect the dynamics of $\phi$ to concavity of $\zeta$. Differentiating gives us
\begin{align*}
\phi'(t)&=\frac{V_t(t,1)}{\mu(1)}\bigg/\zeta'\circ\zeta^{-1}
  \left(\frac{V_t(t,1)}{\mu(1)}\right)\\
&=\frac{V_t(t,1)}{\zeta'(\phi,1)\mu(1)}.
\end{align*}
Now let $w(\theta,x)=\xi(\theta)\mu(x)$, where $\xi$ is more concave than $\zeta$, and consider the corresponding value function $W$.\footnote{More concave means that the coefficient of risk aversion for $w$ is everywhere greater than for $u$. Formally,
\[
-\frac{u''(\theta)}{u'(\theta)}<-\frac{w''(\theta)}{w'(\theta)}.
\]} We assume that $\xi$ is well-behaved in the same way as $\zeta$. If $\psi$ is the cutoff corresponding to $W$, then $\psi'=W'/\xi'(\psi)\mu(1)$. We need two lemmas before we can establish the relationship between $V$ and $W$ as well as $\phi$ and $\psi$. The idea is that when $V$ and $W$ are the same value (which happens at different times $t$), more concave $w$ means that $W'<V'$, so $W$ decreases toward 0 at $T$ faster. But that's possible only if $V(t)<W(t)$. The opposite comparison happens for $\phi$ and $\psi$ because of the factors of $u'$ and $w'$ in the denominators of $\phi'$ and $\psi'$.

\th{differential-eq-compare}

\th{normalize-deriv-int}

\th{phi-psi-compare-single}

The general case is similar except that the maximand in equations~\ref{bellman} and \ref{differential-eq} contains $n+1$ functions of the form $u(\theta,i)+V(t,n-i)$. We integrate the whole thing with respect to $\theta$, and it is helpful to imagine a graph of each maximand-component on $[0,1]$, the probability space of $\theta$-values. Several observations are apparent. First, when $i=n$, which corresponds to spending everything, the maximand-component is proportional to $\zeta$ and time-independent. Second, when $i=0$, which corresponds to saving everything, the maximand-component is constant with respect to $\theta$ and decreasing in time. Third, for other values of $i$, the maximand-component is a scaled, time-independent copy of $\zeta$ plus an amount that decreases in time. Larger constant of proportionality corresponds to smaller vertical intercept, so we have a tradeoff between the scale of each component-graph and its value at $\theta=0$.

The saving rule $y$ satisfies intuitive properties in regard to $\theta$. Suppose that $\theta_1<\theta_2$, and let $y^*$ be the output of the spending rule for $\theta_1$. If $n-i<n-y^*$, then we know
\begin{equation}
u(\theta_1,n-i)+V(t,i)\leq u(\theta_1,n-y^*)+V(t,y^*)
\label{compare-u-theta}
\end{equation}
from the definition of $y^*$. Because $\mu$ is increasing,
\begin{align*}
\mu(n-i)&<\mu(n-y^*)\\
[\zeta(\theta_2)-\zeta(\theta_1)]\mu(n-i)
  &<[\zeta(\theta_2)-\zeta(\theta_1)]\mu(n-y^*)\\
u(\theta_2,n-i)-u(\theta_1,n-i)
  &<u(\theta_2,n-y^*)-u(\theta_1,n-y^*).
\end{align*}
Adding this inequality to \eq{compare-u-theta} shows that the equation is true for $\theta_2$ as well, so $n-y$ is non-strictly increasing in $\theta$, and $y$ is non-strictly decreasing. Because the output of $y$ is discrete, we see that for fixed $t$, $y$ is piecewise constant with respect to $\theta$. Further, \eq{compare-u-theta} for $\theta_2$ will be a strict inequality, and we conclude that for generic $\theta$, the largest maximand-component will be strictly greater than the other maximand-components. Monotonicity of $y$ implies that moving right along the $\theta$-axis causes us to transition from flatter to steeper maximand-components.

Concavity of $V$ is less obvious in the multi-resource case because the heights of the maximand-components decrease differently from one another. Differentiating \eq{differential-eq} gives us
\begin{equation}
\frac{d^2V}{dt^2}\,\bigg|_{(t,x)}=\lambda[V_t(t,x)-\E V_t(t,y^*)],
\label{differential-eq-second-deriv}
\end{equation}
where $V_t$ is the partial derivative of $V$ with respect to $t$, and $y^*$ is the agent's choice of $y$ at $t$. The expected value is with respect to $\theta$ and is important because $y^*$ is a function of $\theta$. At $T$, $y^*=0$, so
\[
\frac{d^2V}{dt^2}\,\bigg|_{(T,x)}=\lambda[V_t(T,x)-V_t(T,0)]=-\lambda^2\E u(\theta,x)<0.
\]
Concavity of $\mu$ means that $V_t(T,x)$ is convex in $x$. It follows that there exists a neighborhood of $T$ where $V$ is concave in $x$, and we can again extend to all $t<T$ through continuous induction. Concavity with respect to $x$ allows us to establish concavity in $t$ and a negative cross-partial derivative. We have the following three lemmas and theorem on the structure of $V$. In the proof of Theorem~\ref{V-properties-divisible}, lemma~\ref{discrete-envelope} plays the role of the envelope theorem.

\th{discrete-envelope}

\th{concave-max}

\th{int-functions}

\th{V-properties-divisible}

\th{y-properties}

Recall that in the case of a single resource, \eq{differential-eq} has a nullcline at $u(1)$. The same idea applies when the agent has a divisible resource stock. Because $\mu$ is concave, the agent most wants to consume each unit of $x$ separately, and if they have a large time window, the agent expects to do so successfully at opportunities where $\theta$ is near 1. Formalizing this intuition through induction on $x$ gives us the following theorem.

\th{asymptote}

A corollary of Theorem~\ref{asymptote} is that the agent exhibits static preference reversals. Static preference reversal in this context means that away from $T$, the agent would rather receive a larger future payment than a smaller payment right now, but the agent expects to change that decision if asked closer to $T$. The agent of my model is time-consistent, so their belief about changing their decision is accurate. We can model this scenario by comparing the value of a current resource $x$ to the value of a larger resource stock $x+\bar x$ at future time $t+\bar t$. The difference
\begin{equation}
V(t,x)-V(t+\bar t,x+\bar x).
\label{static-preference-reversal}
\end{equation}
is positive when the agent would rather have a smaller resource stock now and negative when the agent wants to wait $\bar t$ extra time for a larger resource stock. At time $t=T-\bar t$, equation~(\ref{static-preference-reversal}) is positive because $V(t+\bar t,x+\bar x)=V(T,x+\bar x)=0$. For $t$-values far enough away from the deadline, \eq{static-preference-reversal} approaches $-\bar x\mu(1)<0$. In the language from section~2 of this paper, the evaluation period could be any time $t_0$ such that $V(t_0,x)-V(t_0+\bar t,x+\bar x)<0$, and the consumption period is any time $t$ such that $t_0<t<T-\bar t$. For different values of $t$, the agent at $t_0$ chooses differently between $x$ resources at $t$ and $x+\bar x$ resources at $t+\bar t$.

\th{preference-reversal}

We see that the agent behaves reasonably. As the quality of a given opportunity increases, the marginal utility of consumption also rises, so the agent spends more and saves less. The agent spends more as they get closer to the deadline because the marginal value of saving another resource gets smaller at later times. Concavity of $V$ with respect to $x$ means the agent experiences diminishing marginal value of their resource stock, and concavity with respect to $t$ means that the resource stock loses value increasingly fast as the deadline approaches. The agent exhibits static preference reversal because away from the deadline, waiting $\bar t$ time doesn't matter very much, but near the deadline, that time is much more valuable.

\section{Multiple Payments}

We can modify our framework to describe an agent who receives multiple payments at different times, and in this section, I investigate the agent's preferences for timing and correlation of multiple payments. Suppose the agent has $x$ resources at time $t$, and at some fixed future time $\bar t<T$, the agent will receive an additional $\bar x$ resources. After $\bar t$, the agent has received all their resources and uses the same value function we've been considering, but before $\bar t$, the value function must account for the upcoming payment to the agent. Let $V$ denote the value function from previous sections, and let $\tilde V$ be the value function of an agent who receives the second payment at $\bar t$. We have
\begin{align*}
t\geq\bar t&:\qquad \tilde V(t,x)=
  \int_t^T\E\max\{u(\theta,x-y)+\tilde V(s,y)\}\lambda e^{\lambda(t-s)}\:ds\\
t<\bar t&:\qquad \tilde V(t,x)=
  \int_t^{\bar t}\E\max\{u(\theta,x-y)+\tilde V(s,y)\}\lambda e^{\lambda(t-s)}\:ds
  +e^{\lambda(t-\bar t)}\tilde V(\bar t,x+\bar x)
\end{align*}
The term $\tilde V(\bar t,x+\bar x)$ is a continuation payoff in the situation where the agent moves forward in time until $\bar t$ without encountering an opportunity, and $e^{\lambda(t-\bar t)}$ is the chance of no opportunities arising between $t$ and $\bar t$. Notice that when $t\geq\bar t$, $\tilde V$ satisfies the same recursion relation as \eq{bellman}, so after $\bar t$, $\tilde V=V$.

When both payments occur with positive probability, the agent exhibits correlation aversion. Suppose the agent randomly receives a payment of $x$ at some time $t$ and then randomly receives a payment of $\bar x$ at $\bar t>t$. Let $X_1$ and $X_2$ be Bernoulli random variables that encode the agent's probability of getting paid at $t$ and $\bar t$ respectively. Here is how to construct $X_1$ and $X_2$ to account for correlations in their outcomes: let $X$ be uniform on $[0,1]$, and define $X_1=\chi_{[0,p_1]}(X)$, where $p_1$ is the success probability for $X_1$ and $\chi$ is an indicator function. Then we define $X_2=\chi_{[0,p_2]}(X-c)$, so $X_2$ occurs whenever the realization of $X$ falls in the interval $[c,c+p_2]$. To ensure that every value of $c$ corresponds to a unique correlation, we restrict $c$ to values between $\max\{0,p_1-p_2\}$ and $\min\{p_1,1-p_2\}$. Larger $c$ means less correlation. Both $X_1$ and $X_2$ are perfectly correlated if $p_1=p_2$ and $c=0$, and they are perfectly anti-correlated if $p_1+p_2=1$ and $c$ is maximal. Figure~3 shows the joint probability distribution of $X_1$ and $X_2$ conditional on $c$.

\begin{figure}[t]
\tabcolsep=1.3em
\centerline{\bfseries Figure 3: Probabilities of Outcomes for Correlated Bernoulli Distributions}
\medskip
\centerline{%
\begin{tabular}{lcc}\toprule
& $X_2=0$ & $X_2=1$\\\cmidrule{2-3}
$X_1=0$ & $\min\{1-p_1,1-p_2\}-c$ & $\max\{p_2-p_1,0\}+c$ \\
$X_1=1$ & $\max\{p_1-p_2,0\}+c$ & $\min\{p_1,p_2\}-c$ \\\bottomrule
\end{tabular}}
\caption{The joint probability distribution for two correlated Bernoulli distributions. The parameter $c$ determines the amount of correlation, and as $c$ increases, the distributions become less correlated.}
\end{figure}

Prior to $t$, the expected value $E$ of the double random payment is
\begin{align*}
E=(\min\{p_1,p_2\}-c)\tilde V(t,x)&+(\max\{p_1-p_2,0\}+c)V(t,x)\\
  &{}+(\max\{p_2-p_1,0\}+c)V(\bar t, \bar x),
\end{align*}
so
\[
\frac{dE}{dc}=-\tilde V(t,x)+V(t,x)+V(\bar t, \bar x).
\]
Correlation aversion means that $E$ is increasing, which is equivalent to the condition $\tilde V(t,x)<V(t,x)+V(\bar t, \bar x)$. We know that
\[
\lim_{t\uparrow\bar t}\tilde V(t,x)=V(\bar t,x+\bar x),
\]
so near $\bar t$, correlation aversion holds from concavity of $V$. We can extend this result to all $t<\bar t$ through continuous induction, and we have the following theorem.

\th{correlation-aversion}

We can also establish monotonicity with respect to payment timing, which means that $\tilde V$ decreases as $\bar t$ gets larger and is continuous as $\bar t\downarrow t$. Suppose the agent receives a payment of $x$ at $t$ and a payment of $\bar x$ at $\bar t$. Then $\tilde V$ captures the value of the agent's current resources plus payment at $\bar t$, and we can show that increasing $\bar t$ lowers $\tilde V$. The value function $\tilde V$ is discontinuous at $\bar t$ as written because we designed $\tilde V$ to encode the value of a resource stock that jumps discontinuously at $\bar t$. We consider the composite $\tilde V(t,x+\bar x\chi_{[\bar t,T]}(t))$, which is the value of the agent's resource stock when they get paid at $\bar t$, and this composite is well-behaved when $t$ and $\bar t$ get close to one another. Differentiating with respect to $\bar t$ gives us
\begin{align*}
\frac{d\tilde V}{d\bar t}\,\bigg|_{t,x}
  =\int_t^{\bar t}\E\left[\frac{d\tilde V}{d\bar t}\,\bigg|_{s,y^*}\right]
    \lambda e^{\lambda(t-s)}\:ds
  &+\E\max_{0\leq y\leq x}\{u(\theta,x-y)+V(\bar t,\bar x+y)\}
    \lambda e^{\lambda(t-\bar t)}\\
  &-\E\max_{0\leq y\leq x+\bar x}\{u(\theta,x+\bar x-y)+V(\bar t,y)\}
    \lambda e^{\lambda(t-\bar t)},
\end{align*}
where $y^*$ maximizes $u(\theta,x-y)+\tilde V(s,y)$. This expression is intractable, and differentiating with respect to $t$ results in a slightly nicer differential equation for $d\tilde V/d\bar t$ given by
\begin{equation}
\label{differential-eq-tilde-V}
\frac{d}{dt}\left(\frac{d\tilde V}{d\bar t}\,\bigg|_{t,x}\right)=
  \lambda\frac{d\tilde V}{d\bar t}\,\bigg|_{t,x}-
  \lambda\E\left[\frac{d\tilde V}{d\bar t}\,\bigg|_{t,y^*}\right]
  e^{\lambda(t-\bar t)}.
\end{equation}
We show through induction on $x$ that the solution is always negative when $\bar x>0$, and we end up with the following theorem.

\th{monotonicity}

The results in this section illustrate possible effects of adding liquidity to models of consumption. If the agent is completely illiquid like in traditional approaches to consumption smoothing, the agent consumes $x$ and $\bar x$ immediately, and my model becomes equivalent to additive discounting. In that situation, the agent will be correlation neutral and break monotonicity with respect to payment timing because the marginal value of $\bar x$ changes depending on whether $t=\bar t$. Concavity of $\mu$ also plays a role in correlation aversion. If the agent has some liquidity but convex $\mu$, they spend all resources in the same opportunity, so the multi-resource case and the single-resource case have identical solutions. In that situation, the agent is correlation-seeking.

\section{Risk and Time}

The agent of this paper exhibits an intuitive connection between risk and time, and I investigate the relationship in this section. For notational convenience in this section, we define $V_i(t)=V(t,i)$. The subscript is an identifier, not a partial derivative like when we previously wrote $V_t$. Recall that for fixed $t$, the saving rule is piecewise constant and decreasing with respect to $\theta$, so $y$ exhibits well-defined cutoffs between the $\theta$-values that correspond to different outputs. The cutoffs are important for understanding risk and time. Let $\phi_{i,j}(t)$ be the maximum $\theta$ such that an agent with $i$ resources saves $j$ of them if they encounter an opportunity at $t$. Formally,
\[
\phi_{i,j}(t)=\max\{\theta\colon y(\theta,t,i)=j\}.
\]
Equivalently, we may define $\phi_{i,j}$ as the solution to
\begin{equation}
u(\phi_{i,j}(t),i-j)+V_j(t)=u(\phi_{i,j}(t),i-j+1)+V_{j-1}(t),
\label{phi-i-j-condition}
\end{equation}
and it follows that
\begin{equation}
\phi_{i,j}=\zeta^{-1}\left(\frac{V_j-V_{j-1}}{\mu(i-j+1)-\mu(i-j)}\right).
\label{phi-i-j}
\end{equation}
Because $V_j-V_{j-1}$ is decreasing and continuous, $\phi_{i,j}$ is also decreasing and continuous in $t$. Concavity of $V$ with respect to $x$ means that $\phi_{i+1,j+1}<\phi_{i,j}$, and concavity of $\mu$ means that $\phi_{i,j}<\phi_{i+1,j}$. We see that $\phi_{i,j}$ is increasing in $i$ and decreasing in $j$. At the deadline, $\phi_{i,j}(T)=0$, so the agent eventually tries to spend all their resources if they encounter an opportunity close enough to $T$.

The domain of $\phi_{i,j}$ depends on $i$ and $j$. When $i=j$, 
\[
\phi_{i,i}=\zeta^{-1}\left(\frac{V_i-V_{i-1}}{\mu(1)}\right).
\]
Theorem~(\ref{asymptote}) implies that as $t$ gets big negative, the argument of $\zeta^{-1}$ approaches 1, so $\phi_{i,i}$ is always well-defined. This result means that the agent always maintains a cutoff between spending nothing and spending a single unit from their resource stock. For $i<j$, the agent does not always maintain this cutoff. In the argument of $\zeta^{-1}$ from \eq{phi-i-j}, the numerator approaches $\mu(1)$ away from $T$, so concavity of $\mu$ means the argument of $\zeta^{-1}$ eventually exceeds 1, which makes $\phi_{i,j}$ undefined. When I write $\phi_{i,j}^{-1}(1)$, I refer to the minimum time $t$ where $\phi_{i,j}$ is defined. From properties of $\phi_{i,j}$, we see that $\phi_{i,j}^{-1}(1)$ is increasing in $i$ and decreasing in $j$. When $t\geq\phi_{i+1,j}^{-1}(1)$, it is possible to compare the dynamics of $\phi_{i,j}$ and $\phi_{i+1,j}$. Unfortunately, there is no straightforward way to compare the dynamics of $\phi_{i,j}$ and $\phi_{i,j+1}$.

\th{log-concave-property}

\th{cutoffs-approaching}

We can make statements about concavity if we decompose the time-derivatives $V_i'$ into a formulation that is easier to work with. Recall that $V'_i$ is the negative of the area below all maximand-components and above $V_i$ when we graph everything on the probability space $[0,1]$. We decompose this area into separate contributions from each maximand-component. When $\theta$ falls between $\phi_{i,j+1}$ and $\phi_{i,j}$, the agent elects to save $j$ resources, and the region corresponding to $V_i'$ has height $u(\theta,i-j)+V_j-V_i$. By adding and subtracting different $\Delta u$ and $\Delta V$ terms, we can write $V_i'$ as a sum of integrals that telescope on each region between consecutive cutpoints. We have
\begin{align*}
V_i'&=-\sum_{k=0}^{i-1}\int_{\phi_{i,i-k}}^1
  \Big[u(\theta,k+1)+V_{i-k-1}-u(\theta,k)-V_{i-k}\Big]\:d\theta\\[0.5em]
&=-\sum_{k=0}^{i-1}\big[\mu(k+1)-\mu(k)\big]\int_{\phi_{i,i-k}}^1
  \left[\zeta(\theta)-\frac{V_{i-k}-V_{i-k-1}}{\mu(k+1)-\mu(k)}\right]\:d\theta
\end{align*}
I am implicitly assuming that we evaluate only the terms of the sum where $\phi_{i,i-k}$ is defined. Where it is defined, $\phi_{i,i-k}$ satisfies
\[
\zeta(\phi_{i,i-k})=\frac{V_{i-k}-V_{i-k-1}}{\mu(k+1)-\mu(k)},
\]
so the integral measures the area under $\zeta$ above $\zeta(\phi_{i,i-k})$. 

If we integrate along the vertical axis instead of the horizontal axis, $V'_i$ becomes simpler, and we end up with
\[
V_i'=-\sum_{k=0}^{i-1}\big[\mu(k+1)-\mu(k)\big]
  \int_{\min\left\{\frac{V_{i-k}-V_{i-k-1}}{\mu(k+1)-\mu(k)},1\right\}}^1
  (1-\zeta^{-1}(y))\:dy.
\]
Subtracting consecutive $V_i'$ values gives us a differential equation for $\Delta V_t$.
\begin{align}
V_{i+1}'-V_i'
  =-\sum_{k=0}^{i-1}&\big[\mu(k+1)-\mu(k)\big]
  \int^{\min\left\{\frac{V_{i-k}-V_{i-k-1}}{\mu(k+1)-\mu(k)},1\right\}}
    _{\min\left\{\frac{V_{i+1-k}-V_{i-k}}{\mu(k+1)-\mu(k)},1\right\}}
  (1-\zeta^{-1}(y))\:dy\nonumber\\
  &{}-\big[\mu(i+1)-\mu(i)\big]
  \int_{\min\left\{\frac{V_1}{\mu(i+1)-\mu(i)},1\right\}}^1
  (1-\zeta^{-1}(y))\:dy.
\label{decomposition}
\end{align}
\Eq{decomposition} allows us to easily compare marginal valuations under more or less concave utility functions.

\th{more-concave-w-marginal}

Theorem~\ref{more-concave-w-marginal} means that the agent's marginal value of an extra resource is greater when the agent has more concave utility with respect to $\theta$. Greater marginal value translates to an agent who is less responsive to high-quality opportunities and more responsive to low quality opportunities. Let $\psi_{i,j}$ be the $(i,j)$-cutoff corresponding to $W$. Lemma~\ref{normalize-deriv-int} implies that in a neighborhood of $T$, $\psi_{i,j}<\phi_{i,j}$. However, suppose $j<i$, and consider the time $t=\psi_{i,j}^{-1}(1)>-\infty$. We know that
\begin{align*}
\phi_{i,j}(t)
  &=\zeta^{-1}\left(\frac{V_j-V_{j-1}}{\mu(i-j+1)-\mu(i-j)}\right)\\
  &<\zeta^{-1}\left(\frac{W_j-W_{j-1}}{\mu(i-j+1)-\mu(i-j)}\right)
  =\zeta^{-1}(1)=1=\psi_{i,j}(t),
\end{align*}
so the cutoffs cross over. The less risk-averse agent is willing to spend more resources sooner but only at high-quality opportunities.

This setup lends itself to a model of procrastination where the agent misperceives their instantaneous utility function and overestimates their eagerness to spend resources. Suppose that instead of learning $\theta$ at an opportunity, the agent learns possible output values of $u(\theta,x-y)$ for different values of $y$. The agent derives utility from the act of consuming $x-y$, and they know exactly how much utility they will receive from spending $x-y$ resources but not the specifics of their own utility function or the realized $\theta$. Suppose that instead of $\zeta$, the agent believes that their instantaneous utility function is $\tilde\zeta$, where $\tilde\zeta$ is more concave than $\zeta$. Whenever the agent reaches an opportunity with quality $\theta$, they believe it has quality $\tilde\theta=\tilde\zeta^{-1}(\zeta(\theta))$ because $\tilde\zeta(\tilde\theta)=\zeta(\theta)$. The agent decides to spend resources according to the solution for $\tilde\zeta$, so they save $j$ resources if \smash{$\tilde\theta<\tilde\phi_{i,j}$}, where $\tilde\phi_{i,j}$ is the $(i,j)$-cutoff under \smash{$\tilde\zeta$}.

In terms of $\theta$ rather than \smash{$\tilde\theta$}, this cutoff is given by
\begin{align*}
\tilde\zeta^{-1}(\zeta(\theta))=\tilde\theta<\tilde\phi_{i,j}
  &=\tilde\zeta^{-1}\left(
    \frac{\tilde V_j-\tilde V_{j-1}}{\mu(i-j+1)-\mu(i-j)}\right)\\
\zeta(\theta)
  &<\frac{\tilde V_j-\tilde V_{j-1}}{\mu(i-j+1)-\mu(i-j)}\\
\theta&<\zeta^{-1}\left(
    \frac{\tilde V_j-\tilde V_{j-1}}{\mu(i-j+1)-\mu(i-j)}\right),
\end{align*}
where $\tilde V$ is the solution to the model for an agent with utility $\tilde\zeta$. As we saw in \eq{phi-i-j}, a fully informed agent saves $j$ resources when
\[
\theta<\zeta^{-1}\left(
    \frac{V_j-V_{j-1}}{\mu(i-j+1)-\mu(i-j)}\right).
\]
Because $\tilde\zeta$ is more concave than $\zeta$, the marginal value $\tilde V_j-\tilde V_{j-1}$ is greater than $V_j-V_{j-1}$, so the procrastinating agent uses a saving rule that is too high. The agent always spends less than they would if they perceived their utility function correctly, which delays spending.

A similar situation emerges if the agent accurately knows $\theta$ but misperceives the nature of time. Suppose the Poisson opportunities arise at rate $\lambda$, but the agent believes they occur at rate $\tilde\lambda>\lambda$. Let $V$ correspond to the valuation of resources by an agent who accurately knows how often opportunities appear, and let $\tilde V$ be the valuation when the agent believes that opportunities occur at $\tilde\lambda$. Intuitively, the agent overestimates their capacity to spend future resources, we can see the effect of this assumption by comparing the saving rules $y$ and $\tilde y$. We need the following lemma.

Let $\kappa=\tilde\lambda/\lambda$. From lemma~\ref{scale-diff-eq} and induction on $i$, we see that for any $i$, we have
\[
\tilde V'_i(t)=\kappa V'_i(t),
\]
and $\tilde V$ is a horizontal dilation of $V$ around $T$. Because $\kappa>1$, the dilation shrinks $V$ to produce $\tilde V$, so for any $t$, there exists $\tilde t<t$ such that for any $i$, $\tilde V_i(t)=V_i(\tilde t)$. It follows that
\[
\tilde V_i-\tilde V_{i-1}>V_i-V_{i-1}.
\]
For the same reasoning as the case with misperceived $\zeta$, the agent uses a higher cutoff when they believe the opportunity rate to be $\tilde\lambda$. Once again, the agent saves more resources than they would with perfect information, and we can think of this behavior as a form of procrastination. The agent's incorrect beliefs about the world prompt them to delay spending resources that they would otherwise consume.

\section{Conclusion}

I constructed a rational agent who spends resources at random opportunities between current time $t$ and some future deadline $T$. The agent's value function is relatively simple, but the agent exhibits intuitive and interesting behavior. The agent's valuation $V$ is positive, decreasing in $t$, and increasing in $x$, and the same is true of the saving rule $y$. Further, $V$ is concave in both variables, and the marginal value with respect to time gets more negative as $x$ increases. In the limit away from the deadline, the agent plans to spend each unit of $x$ separately at an opportunity with large $\theta$, so $V(t,x)$ has a horizontal asymptote at $x\mu(1)$. Near the deadline, the agent feels increasing pressure to spend resources as quickly as possible. Because they have some liquidity, the agent exhibits both correlation aversion and monotonicity with respect to payment timing. Connecting risk and time is sensible, and doing so produces a model of an agent who procrastinates because they misperceive their own utility function or the nature of time passing. These solution properties describe an agent who is very aware of the passage of time and plans accordingly to meet deadline pressure. More research is needed to test these ideas empirically and to determine whether they lead to a more general model of time preference, apart from deadlines.

\bigskip
\centerline{\bfseries References}

\vskip-\medskipamount
\vskip\z@
{
\def\mkbibnamefamily#1{\textsc{#1}}
\def\mkbibnamegiven#1{\textsc{#1}}
\def\mkbibnameprefix#1{\textsc{#1}}
\def\mkbibnamesuffix#1{\textsc{#1}}

\printbibliography[heading=none]

\vskip-\lastskip

}

\vfill\eject

\centerline{\bfseries Appendix A: Derivation of the Bellman Equation}

\medskip

\noindent In this paper, a rational agent faces a decision problem of spending resources at current time $t$ to maximize current instantaneous utility plus the expected sum of future instantaneous utility. Suppose that $x_0$ and $\theta_0$ represent current spending and quality, respectively, and let $x_i$ and $\theta_i$ denote spending and quality at the $i$th future opportunity. The randomness in this model comes from a Poisson process and independent uniform random variables. Let $\Omega$ be a probability space that encodes the timing of Poisson events and the associated quality values, and let $\{F_t\}$ be a filtration indexed by time, where $F_t$ allows full understanding of everything at or before $t$ but nothing afterwards.

We can define such a set without too much trouble. Let $A$ denote the set of countable, increasing, well-ordered strings of real numbers less than or equal to $T$, and then
\[
\Omega=\bigsqcup_{s\in A}\{s\}\times[0,1]^{s}.
\]
Each outcome in the probability space is a pairing of a string $s$ of increasing real numbers and a function $\gamma$ from $s$ to $[0,1]$. The entries in $s$ are the times of Poisson events, and $\gamma$ encodes the quality parameter for each Poisson event. The measure on $A$ arises from the Poisson process, and we put the Lebesgue measure on each copy of $[0,1]$. The $\sigma$-algebra $F_t$ inherits its structure from the Lebesgue $\sigma$-algebra on $(-\infty,t]$ and on those copies of $[0,1]$ that correspond to Poisson events occurring at or before $t$. Then for any $F_t$, two outcomes $\omega_1=(s_1,\gamma_1)$ and $\omega_2=(s_2,\gamma_2)$ are in the same element of $F_t$ if and only if the following properties are true: (1) the portions of $s_1$ and $s_2$ that are less than or equal to $t$ are the same and (2) $\gamma_1$ and $\gamma_2$ agree on those (identical) portions of $s_1$ and $s_2$.

In mathematical symbols, the agent solves
\[
\max_{x_0,x_1,{\dots}}u(x_0,\theta_0)+\E\left[\sum_{i=1}^nu(x_i,\theta_i)\middle|F_t\right]
\]
subject to
\[
\sum_{i=0}^nx_i\leq x,
\]
where $n$ is the total number of opportunities. The agent does not know $n$ prior to reaching $T$, and the agent does not know $\theta_i$ prior to reaching the $i$th future opportunity. Formally, $n$, $x_i$, and $\theta_i$ are themselves random variables. Here $n$ is $F_T$-measurable, and $x_i$ and $\theta_i$ are required to be measurable with respect to $F_{t_i}$, where $t_i$ is the time of the $i$th future spending opportunity.

For a given $x$ and $t$, define 
\begin{equation}
V(t,x)=\max_{x_1,x_2,{\dots}}\E\left[\sum_{i=1}^nu(x_i,\theta_i)\middle|F_t\right]
\label{define-V}
\end{equation}
subject to
\[
\sum_{i=0}^nx_i\leq x.
\]
We have to prove that $V$ is well-defined, which means showing the right side of \eq{define-V} is constant. I claim this property follows from independence of the opportunity qualities and the nature of the Poisson process. Informally, knowing the history of the Poisson process and opportunity qualities tells the agent nothing about the future.

Let $G_t$ be the $\sigma$-algebra that is complementary to $F_t$ in the sense that it inherits its structure from the Lebesgue measure on $(t,T]$ and those copies of $[0,1]$ that correspond to Poisson events occurring after $t$. Let $\omega_1=(s_1,\gamma_1)$ and $\omega_2=(s_2,\gamma_2)$ be two outcomes in $\Omega$ such that $s_1$ and $s_2$ agree on $(t,T]$ and such that $\gamma_1$ and $\gamma_2$ are equal on that portion of $s_1$ (equivalently $s_2$). This means $\omega_1$ and $\omega_2$ are contained in exactly the same events in $G_t$. If we think of the optimal $\{x_i\}$ as a finite sequence of random variables, then it must be the case that the sum on the right side of \eq{define-V} has the same value when evaluated at $\omega_1$ or $\omega_2$. Otherwise we could replace $x_i(\omega_1)$ with $x_i(\omega_2)$ or vice versa and increase the value of the sum. It follows that we can assume without loss of generality that $x_i(\omega_1)=x_i(\omega_2)$, so each optimal $x_i$ is $G_t$-measurable. However, from the definition of Poisson process and the independence of the quality parameters, $F_t$ and $G_t$ are independent. Because under the optimal $x_i$, the sum is also $G_t$-measurable, it follows that the right side of \eq{define-V} is constant.

If we let $t^+$ denote the time of the next Poisson event after $t$, then we can write $V$ recursively. We have
\begin{align*}
V(t,x)&=\max_{x_1,x_2,{\dots}}\E\left[\sum_{i=1}^nu(x_i,\theta_i)\middle|F_t\right]\\
&=\max_{x_1,x_2,{\dots}}\E\left[\E\left(\sum_{i=1}^nu(x_i,\theta_i)\middle|F_{t^+}\right)\middle|F_t\right]\\
&=\max_{x_1}\max_{x_2,x_3,{\dots}}\E\left[u(x_1,\theta_1)+\E\left(\sum_{i=2}^nu(x_i,\theta_i)\middle|F_{t^+}\right)\middle|F_t\right]
\end{align*}
We converted $V$ into a compound problem where the agent picks $x_1$ separately from $x_2$ onward. We know that the agent picks $x_1$ at time $t^+$ after learning $\theta_1$, so we can rewrite $V$ as the expected value of the agent's payoff when the agent reaches $t^+$. The possible times for $t^+$ are distributed exponentially, so we have
\begin{align*}
V(t,x)&=\int_t^T\E\max_{x_1}\max_{x_2,x_3,{\dots}}\left[u(x_1,\theta_1)+
  \E\left(\sum_{i=2}^nu(x_i,\theta_i)\middle|F_{s},t^+=s\right)\right]
  \lambda e^{\lambda(t-s)}\:ds\\
&=\int_t^T\E\max_{x_1}\left[u(x_1,\theta_1)+
  \max_{x_2,x_3,{\dots}}
  \E\left(\sum_{i=2}^nu(x_i,\theta_i)\middle|F_{s},t^+=s\right)\right]
  \lambda e^{\lambda(t-s)}\:ds,
\end{align*}
and this is equivalent to \eq{bellman}.

\vfill\eject

\centerline{\bfseries Appendix B: Mathematical Proofs}

\c@definition\z@
\let\th@label\@gobble

\medskip

\noindent This appendix contains proofs of the theorems and lemmas in the paper.

\th{scale-by-k}

\begin{proof}
Suppose $V$ solves the Bellman equation with instantaneous utility $u$. Integrals are linear operators, and maximization commutes with multiplication by a positive constant. If we multiply \eq{bellman} by $k$, we get
\[
kV(t,x)=\int_t^T\E\max\{ku(\theta,x-y)+kV(s,y)\}\lambda e^{\lambda(t-s)}\:ds,
\]
so it follows that $kV$ solves the Bellman equation with instantaneous utility $ku$. Multiplying by a positive constant does not affect which component of the maximand is greatest, so the agent's choice of $y$ is unaffected.
\end{proof}

\th{positive-increasing-V}

\begin{proof}
It is clear from \eq{differential-eq} that $V$ is non-strictly decreasing. Let $y^*(\theta,t,x)$ be the saving rule. It follows that
\begin{align*}
V(t,x)&=\int_t^T\E\{u(\theta,x-y^*(\theta,s,x))+V(s,y^*(\theta,s,x))\}\lambda e^{\lambda(t-s)}\:ds\\
  &\leq\int_t^T
    \E\{u(\theta,x-y^*(\theta,s,x))+V(t,y^*(\theta,s,x))\}\lambda e^{\lambda(t-s)}\:ds\\
  &\leq\int_t^T
    \E\max\{u(\theta,x-y)+V(t,y)\}\lambda e^{\lambda(t-s)}\:ds\\
  &=\E\max\{u(\theta,x-y)+V(t,y)\}\int_t^T\lambda e^{\lambda(t-s)}\:ds\\
  &<\E\max\{u(\theta,x-y)+V(t,y)\},
\end{align*}
so $V$ is strictly decreasing in time, assuming $x>0$. Positivity follows. To show increasing in $x$, let $x_1<x_2$. Then
\begin{align*}
V(t,x_1)&=\int_t^T\E\{u(\theta,x_1-y^*(\theta,s,x_1))+V(s,y^*(\theta,s,x_1))\}\lambda e^{\lambda(t-s)}\:ds\\
  &<\int_t^T\E\{u(\theta,x_2-y^*(\theta,s,x_1))+V(s,y^*(\theta,s,x_1))\}\lambda e^{\lambda(t-s)}\:ds\\
  &\leq\int_t^T\E\max_{0\leq y\leq x_2}\{u(\theta,x_2-y)+V(s,y)\}\lambda e^{\lambda(t-s)}\:ds=V(t,x_2)
\end{align*}
When $t=T$ or $x=0$, $V$ is identically 0.
\end{proof}

\th{scale-diff-eq}

\begin{proof}
Let $\tilde y(t)=y(T+\lambda(t-T))$. Differentiating $\tilde y$ gives us
\begin{align*}
\frac{d\tilde y}{dt}&=\frac d{dt}y(T+\lambda(t-T))\\
&=\lambda y'(T+\lambda(t-T))\\
&=\lambda f(T+\lambda(t-T),y(T+\lambda(t-T)))=\lambda f(T+\lambda(t-T),\tilde y).
\end{align*}
Clearly $\tilde y(T)=0$, so $\tilde y$ satisfies the same differential equation and initial condition as $z$.
\end{proof}

\th{scale-V}

\begin{proof}
We prove this theorem for discrete $x$ by induction. We begin by noting that $y(T+\lambda(t-T))$ is exactly the transformation that horizontally dilates a function $y$ around $T$ by a factor of $1/\lambda$, so if we can apply lemma~\ref{scale-diff-eq} to $V$, we are done. Lemma~\ref{scale-diff-eq} trivially holds for differential equations without time dependence, which establishes the result for $V(t,1)$. Now assume that for any $k<n$, the theorem holds for $V(t,k)$. But then we can apply lemma~\ref{scale-diff-eq} to \eq{differential-eq} for $V(t,n)$, so the result must hold for $V(t,n)$. By induction, the theorem applies to all $V(t,x)$ as well.
\end{proof}

\th{concave-V-single}

\begin{proof}
If $t_1<t_2$, then $V(t_1)>V(t_2)$. This means $\phi(t_1)=u^{-1}(V(t_1))>u^{-1}(V(t_2))=\phi(t_2)$, so we have
\[
V'(t_1)=-\int_{\phi(t_1)}^1 u(\theta)-\phi(t_1)\:d\theta>-\int_{\phi(t_1)}^1 u(\theta)-\phi(t_2)\:d\theta.
\]
The integrand is positive, so reducing the lower bound makes the integral greater. The minus sign reverses this change, and we have
\[
V'(t_1)>-\int_{\phi(t_2)}^1 u(\theta)-\phi(t_2)\:d\theta=V'(t_2).
\]
This means the derivative is decreasing in $t$, so $V$ is concave.
\end{proof}

\th{differential-eq-compare}

\begin{proof}
There exists a neighborhood of $T$ where $z<y$. Suppose $z<y$ on the interval $(t,T)$. We will show that for some $\epsilon>0$, the inequality holds on $(t-\epsilon,T)$, and we will use this fact to show that the inequality holds everywhere. We know that $z(t)\leq y(t)$ by assumption. If the inequality is strict, we're done. Otherwise, assume that $z(t)=y(t)$. From the mean value theorem, we know that for any $\delta>0$, there exists a $c$ such that
\[
\frac{z(t+\delta)-z(t)}{y(t+\delta)-y(t)}=\frac{z'(c)}{y'(c)}.
\]
We assumed that $z(t+\delta)<y(t+\delta)$, so the quotient must be greater than 1. It follows that
\[
\lim_{\delta\to0}\frac{z'(c)}{y'(c)}=\frac{z'(t)}{y'(t)}\geq 1,
\]
so $z'(t)\leq y'(t)$. (We switch the inequality sign because $y'<0$.) But if $z(t)=y(t)$, then $y'(t)=f(t,y(t))<g(t,z(t))=z'(t)$. By contrapositive, we cannot have $y(t)=z(t)$.

Now let $E$ be the set of points in $(-\infty,T)$ where $z\geq y$. We know that $E$ is bounded above by $T$, so let $c=\sup E$. We know that $c<T$, so $z<y$ on $(c,T)$. However, we also know that for any interval $(t,T)$ where $z<y$, the inequality holds on a larger interval containing $(t,T)$. It follows that $c$ can't exist, and $z<y$ everywhere.
\end{proof}

\th{normalize-deriv-int}

\begin{proof}
From properties of concavity, we know that $g'(x)/f'(x)$ is increasing in $x$. This means that for any $r>x$,
\[
\frac{g'(x)}{f'(x)}<\frac{g'(r)}{f'(r)}\qquad\Rightarrow\qquad\frac{f'(r)}{f'(x)}<\frac{g'(r)}{g'(x)}.
\]
We have
\[
\int_x^b\frac{f(s)-f(x)}{f'(x)}\:ds
  =\int_x^b\!\!\int_x^s\frac{f'(r)}{f'(x)}\:drds\\
  <\int_x^b\!\!\int_x^s\frac{g'(r)}{g'(x)}\:drds\\
  =\int_x^b\frac{g(s)-g(x)}{g'(x)}\:ds.
\]
\end{proof}

\th{phi-psi-compare-single}

\begin{proof}
We begin by comparing value functions. Consider $t_1$ and $t_2$ such that $V(t_1)=W(t_2)$. By assumption, $u<v$, so
\begin{align*}
V'(t_1)&=-\int_{u^{-1}(V(t_1))}^1u(\theta)-V(t_1)\:d\theta\\
&>-\int_{u^{-1}(V(t_1))}^1v(\theta)-V(t_1)\:d\theta\\
&>-\int_{v^{-1}(V(t_1))}^1v(\theta)-V(t_1)\:d\theta\\
&=-\int_{v^{-1}(W(t_2))}^1v(\theta)-W(t_2)\:d\theta=W'(t_2).
\end{align*}
Lemma~\ref{differential-eq-compare} implies that $V<W$.

For cutpoints, recall that $\phi=u^{-1}(V)$ and $\psi=v^{-1}(W)$. We have
\begin{align*}
\phi'&=\frac{V'}{u'\circ u^{-1}(V)}=\frac{V'}{u'(\phi)}
  &\psi'&=\frac{W'}{v'\circ v^{-1}(W)}=\frac{W'}{v'(\psi)}.
\end{align*}
Now consider $t_1$ an $t_2$ such that $\phi(t_1)=\psi(t_2)$. We see that
\begin{align*}
\phi'(t_1)&=\frac{V'(t_1)}{u'(\phi(t_1))}&
  \psi'(t_2)&=\frac{W'(t_2)}{v'(\psi(t_2))}\\[\smallskipamount]
&=-\frac1{{u'(\phi(t_1))}}
    \int_{\phi(t_1)}^1u(\theta)-u(\phi(t_1))\:d\theta&
  &=-\frac1{v'(\psi(t_2))}\int_{\psi(t_2)}^1v(\theta)-v(\psi(t_2))\:d\theta
\end{align*}
Lemma~\ref{normalize-deriv-int} implies that $\phi'(t_1)<\psi'(t_2)$, and applying Lemma~\ref{differential-eq-compare} gives us $\psi<\phi$.
\end{proof}

\th{discrete-envelope}

\begin{proof}
Fix $p$, and let $E$ be the set of points where $\beta|_p$ jumps. By assumption, $E$ is finite and therefore measure 0. Therefore
\begin{align*}
\frac d{dp}\int_0^1 f(s,p,\beta(s,p))\:ds
  &=\frac d{dp}\int_{E^c} f(s,p,\beta(s,p))\:ds\\
  &=\int_{E^c}\frac d{dp}f(s,p,\beta(s,p))\:ds\\
  &=\int_{E^c}\lim_{h\to0}
    \frac{f(s,p+h,\beta(s,p+h))-f(s,p,\beta(s,p))}h\:ds
\end{align*}
By definition of $E^c$, for any $s$ in the region of integration, $\beta$ is locally constant at $(s,p)$. This means that for purposes of calculating the limit, we can replace $\beta(s,p+h)$ with $\beta(s,p)$, and we end up with
\begin{align*}
\frac d{dp}\int_0^1 f(s,p,\beta(s,p))\:ds
  &=\int_{E^c}\lim_{h\to0}
    \frac{f(s,p+h,\beta(s,p))-f(s,p,\beta(s,p))}h\:ds\\
  &=\int_{E^c}\frac{\partial f}{\partial p}\,\bigg|_{s,p,\beta(s,p)}\:ds
    =\int_0^1\frac{\partial f}{\partial p}\,\bigg|_{s,p,\beta(s,p)}\:ds,
\end{align*}
where we freely add $E$ back into the domain of integration because $E$ has measure 0.
\end{proof}

\th{concave-max}

\begin{proof}
We take it as obvious that $h(0)=0$. Consider the sequences of discrete derivatives $F=(\Delta f(i))_i$ and $G=(\Delta g(i))_i$. From concavity of $f$ and $g$, these sequences are strictly decreasing, and for generic $f$ and $g$, no two elements from different sequences are equal. It follows that we can interlace $F$ and $G$ to create a new sequence $A$ of differences that is strictly decreasing, and the value of $h(x)$ will be the sum of the first $x$ elements of $A$. If $A_i$ is the $i$th element of $A$, then $\Delta h(x)=A_x>0$, so $h$ is strictly decreasing. Because $A$ is strictly decreasing, $h$ is strictly concave. The differences $\Delta f(x)$ and $\Delta g(x)$ are the $x$th elements of $F$ and $G$ respectively. Because $A$ contains exactly the elements of $F$ and $G$ in descending order, it must be the case that $F_x$ and $G_x$ appear in $A$ on or after position $x$. This means $\max\{\Delta f(x),\Delta g(x)\}=\max\{F_x,G_x\}\leq A_x=\max\{\Delta f(x-y^*),\Delta g(y^*)\}=\Delta h(x)$. When $0<y^*<x$, the first $x$ elements of $A$ contain elements from both $F$ and $G$, so $F_x$ and $G_x$ appear in $A$ after position $x$. Strictness of the inequality follows in that case.
\end{proof}

\th{int-functions}

\begin{proof}
Discreteness is obvious. Positivity and increasing follow from monotonicity of the integral. To show concavity, let $x_1<x_2$. Then for any $i$, $\Delta f(x_2)<\Delta f(x_1)$, and
\[
\left(\Delta\int_0^1f_i\:di\middle)\right|_{x_2}
  =\int_0^1\Delta f_i(x_2)\:di
  <\int_0^1\Delta f_i(x_1)\:di
  =\left(\Delta\int_0^1f_i\:di\middle)\right|_{x_1}
\]
\end{proof}

\th{V-properties-divisible}

\begin{proof}
We know that $V_t(T)=-\lambda\E u(\theta,x)$, so strict concavity of $\mu$ means that $V$ is strictly concave with respect to $x$ in a neighborhood of $T$. Suppose that $V$ is strictly concave with respect to $x$ on the interval $(t,T)$. Consider some time $s$ satisfying $t<s<T$. For all but at most a countable number of quality values $\theta$, the functions $u(\theta,x)$ and $V(s,x)$ will satisfy the assumptions of Lemma~\ref{concave-max}, and Lemma~\ref{int-functions} means that $\E\max\{u(\theta,x-y)+V(s,y)\}$ is strictly concave. If $x_1<x_2$, we have
\begin{align*}
\Delta V(t,x_1)&=\int_t^T\Delta\E\max\{u(\theta,x_1-y)+V(s,y)\}\lambda e^{\lambda(t-s)}\:ds\\
  &>\int_t^T\Delta\E\max\{u(\theta,x_2-y)+V(s,y)\}\lambda e^{\lambda(t-s)}\:ds=\Delta V(t,x_2),
\end{align*}
so $V$ is also strictly concave at $t$. This means there exists an interval $(t-\epsilon, T)$ where $V$ is strictly concave with respect to $x$, and the result follows for the same reasoning as in the proof of Lemma~\ref{differential-eq-compare}.

Now consider $\Delta V_t(t,x)=\lambda(\Delta V(t,x)-\E\Delta\max\{u(\theta,x-y)+V(t,y)\})$. From Lemma~\ref{concave-max}, we know that $\Delta V_t\leq0$. If $0<y^*<x$ with positive probability, where $y^*$ is the saving rule, then the inequality will be strict. Let \smash{$\underline\phi$} be the cutoff between spending nothing and spending one resource, and let $\overline\phi$ be the cutoff between spending $n-1$ and $n$ resources. From \eq{phi-i-j}, we have
\begin{align*}
\underline\phi&=\zeta^{-1}\left(\frac{V(t,n)-V(t,n-1)}{\mu(1)}\right)&
  \overline\phi&=\zeta^{-1}\left(\frac{V(t,1)}{\mu(n)-\mu(n-1)}\right).
\end{align*}
Concavity of $V$ means the numerator of \smash{$\underline\phi$} is less than the numerator of $\overline\phi$, and concavity of $\mu$ means the denominator is greater. It follows that $\hbox{\smash{$\underline\phi$}}<\overline\phi$, so $\Delta V_t<0$.

Let $y^*$ be the saving rule. We can write \eq{differential-eq} as
\[
\frac{dV}{dt}\bigg|_{(t,x)}=\lambda(V(t,x)-\E u(\theta,x-y^*)-V(t,y^*)).
\]
Differentiating with respect to $t$ and applying Lemma~\ref{discrete-envelope} gives us \eq{differential-eq-second-deriv}. Because $\overline\phi>0$, $y^*<x$ with positive probability, so $\Delta V_t<0$ implies $V_t(t,x)<\E V_t(t,y^*)$. It follows that $V$ is concave in $t$.
\end{proof}

\th{y-properties}

\begin{proof}
Let $y^*$ be the output of the spending rule for $t_1$, $\theta$, and $x$. Consider $t_2>t_1$, and suppose $y^*<i$. By assumption, we have
\[
u(\theta,x-i)+V(t_1,i)\leq u(\theta,x-y^*)+V(t_1,y^*),
\]
and from Theorem~\ref{V-properties-divisible}, we know
\[
V(t_2,i)-V(t_1,i)<V(t_2,y^*)-V(t_2,y^*).
\]
If we add these inequalities together, it follows that
\[
u(\theta,x-i)+V(t_2,i)<u(\theta,x-y^*)+V(t_2,y^*),
\]
so $y^*$ cannot be increasing in $t$.

The second part of this corollary is equivalent to the statement that when $x$ increases by 1, $y$ increases by 0 or 1. This result follows from the same reasoning as in the proof of Lemma~\ref{concave-max}. Suppose that $y^*$ is the saving rule for $t$, $\theta$, and $x$, and we increase $x$ by 1 resource. Then we have
\begin{equation}
\Delta\max_y\{u(\theta,x-y)+V(t,y)\}=\max\{\Delta u(\theta,x-y^*),\Delta V(t,y^*)\},
\label{Delta-max-look-at-y}
\end{equation}
where $\Delta$ means discrete derivative in $x$ as usual. Consider the right side of \eq{Delta-max-look-at-y}. If the first element of the maximand is larger, then increasing $x$ by 1 resource leaves $y$ unchanged, so $y(\theta,t,x+1)=y^*$. If the second element of the maximand is larger, increasing $x$ also increases $y$, and we get $y(\theta,t,x+1)=y^*+1$.
\end{proof}

\th{asymptote}

\begin{proof}
We know this property is true for $x=1$ because $\mu(1)$ is the smallest nullcline of \eq{differential-eq} in this case. Suppose the result holds for $x-1$ or fewer resources, and we will show it must hold for $x$. We have by assumption that for any $y<x$, $u(\theta,x-y)+V(t,y)<x\mu(1)$, and it follows that $V(t,x)=x\mu(1)$ is a nullcline of \eq{differential-eq}. Now consider any $k<x\mu(1)$. By assumption, when $t$ is far away from $T$, $V(t,x-1)\approx(x-1)\mu(1)$, and when $\theta$ is close to 1, $u(\theta,1)\approx\mu(1)$. It follows that there must exist some $t$ such that $u(\theta,1)+V(t,x-1)>k$ on a set of $\theta$-values with positive measure, so $k$ cannot be a nullcline for $V(t,x)$. This means $x\mu(1)$ is the smallest possible nullcline for $V(t,x)$, so $V(t,x)$ must be asymptotic to $x\mu(1)$.
\end{proof}

\th{preference-reversal}

\begin{proof}
Consider the difference $V(t,x)-V(t+\bar t,x+\bar x)$. When the expression is positive, the agent prefers to receive $x$ resources now, and when the expression is negative, the agent prefers to receive $x+\bar x$ resources at $\bar t$ time into the future. For $t=T-\bar t$, the difference is clearly positive because $V((T-\bar t)+\bar t,x+\bar x)=V(T,x+\bar x)=0$. When $t$ gets big negative, the expression is eventually negative from Theorem~\ref{asymptote}. It follows from continuity of $V$ that the agent's preference switches at some point in time. Now differentiate with respect to t. We have
\begin{align*}
\frac d{dt}V(t,x)-V(t+\bar t,x+\bar x)&=V_t(t,x)-V_t(t+\bar t,x+\bar x)\\
&=V_t(t,x)-V_t(t+\bar t,x)+V_t(t+\bar t,x)-V_t(t+\bar t,x+\bar x).
\end{align*}
Concavity of $V$ in $t$ means $V_t(t,x)-V_t(t+\bar t,x)>0$, and $\Delta V_t<0$ implies that $V_t(t+\bar t,x)-V_t(t+\bar t,x+\bar x)>0$. It follows that $V(t,x)-V(t+\bar t,x+\bar x)$ is strictly increasing in $t$, so the sign change happens at a unique point in time.
\end{proof}

\th{correlation-aversion}

\begin{proof} Correlation aversion at time $s$ is equivalent to the condition $\tilde V(s,x)<V(s,x)+V(\bar t, \bar x)$. As mentioned previously, we know that
\[
\lim_{s\uparrow\bar t}\tilde V(s,x)=V(\bar t,x+\bar x),
\]
so near $\bar t$, correlation aversion holds from strict concavity of $V$. Let $\tilde y^*$ be the saving rule for $\tilde V$ at $\theta$, $s$, and $x$, and assume correlation aversion holds on the interval $(t,\bar t)$. Then
\begin{align*}
\tilde V(t,x)&=\int_t^{\bar t}\E\max\{u(\theta,x-y)+V(s,y)\}
  \lambda e^{\lambda(t-s)}\:ds+
  e^{\lambda(t-\bar t)}\tilde V(\bar t,x+\bar x)\\
&=\int_t^{\bar t}\E\left[u(\theta,x-\tilde y^*)+\tilde V(s,\tilde y^*)\right]
  \lambda e^{\lambda(t-s)}\:ds\\
  &\qquad{}+e^{\lambda(t-\bar t)}\tilde V(\bar t,x+\bar x)\\
&<\int_t^{\bar t}\E\left[u(\theta,x-\tilde y^*)+
  V(s,\tilde y^*)+V(\bar t,\bar x)\right]
  \lambda e^{\lambda(t-s)}\:ds\\
  &\qquad{}+e^{\lambda(t-\bar t)}\tilde V(\bar t,x+\bar x)\\
&=\int_t^{\bar t}\E\left[u(\theta,x-\tilde y^*)+
  V(s,\tilde y^*)\right]\lambda e^{\lambda(t-s)}\:ds+
  (1-e^{\lambda(t-\bar t)})V(\bar t,\bar x)\\
  &\qquad{}+e^{\lambda(t-\bar t)}
    [V(\bar t,x)+V(\bar t,\bar x)]\\
&=\int_t^{\bar t}\E\left[u(\theta,x-\tilde y^*)+
  V(s,\tilde y^*)\right]\lambda e^{\lambda(t-s)}\:ds+
  V(\bar t,\bar x)+e^{\lambda(t-\bar t)}V(\bar t,x)\\
&=\int_t^{\bar t}\E\left[u(\theta,x-\tilde y^*)+
  V(s,\tilde y^*)\right]\lambda e^{\lambda(t-s)}\:ds+
  V(\bar t,\bar x)\\
  &\qquad{}+\int_{\bar t}^{T}\E\max\{u(\theta,x-y)+
  V(s,\tilde y)\}\lambda e^{\lambda(t-s)}\:ds\\
&\leq V(t,x)+V(\bar t,\bar x).
\end{align*}
This means that there exists some $\epsilon>0$ such that correlation aversion holds on $(t-\epsilon,\bar t)$. For the same reasoning from the end of the proof of Lemma~\ref{differential-eq-compare}, we see that correlation aversion holds for all $t<\bar t$.
\end{proof}

\th{monotonicity}

\begin{proof}
The value of the agent's resource stock is given by $\tilde V(t,x+\bar x\chi_{[\bar t,T]}(t))$, where $\chi$ is an indicator function. Changing $\bar t$ means that we change not only when the agent receives an extra $\bar x$ resources but also when the agent expects to receive these resources, so for notational clarity, we add a third argument $\bar t$ to $\tilde V$. The third argument tracks when the agent expects the additional payment to happen. If we differentiate $\tilde V$ with respect to $\bar t$, we have
\begin{align*}
\frac d{d\bar t}\tilde V(t,x+\bar x\chi_{[\bar t,T]}(t),\bar t)
  &=\lim_{h\to0}\frac{V(t,x+\bar x\chi_{[\bar t+h,T]}(t),\bar t+h)
    -V(t,x+\bar x\chi_{[\bar t,T]}(t),\bar t)}{h}\\
  &=\lim_{h\to0}\frac1h\bigg[\int_t^{\bar t}\E\max\{u(\theta,x-y)
      +\tilde V(s,y,\bar t+h)\}\lambda e^{\lambda(t-s)}\:ds\\
    &\qquad{}+\int_{\bar t}^{\bar t+h}\E\max\{u(\theta,x-y)
      +\tilde V(s,y,\bar t+h)\}\lambda e^{\lambda(t-s)}\:ds\\
    &\qquad{}-\int_t^{\bar t}\E\max\{u(\theta,x-y)
      +\tilde V(s,y,\bar t)\}\lambda e^{\lambda(t-s)}\:ds\\
    &\qquad{}-\int_t^{\bar t}\E\max\{u(\theta,x+\bar x-y)
      +\tilde V(s,y,\bar t)\}
      \lambda e^{\lambda(t-s)}\:ds\bigg]\\[\medskipamount]
  &=\int_t^{\bar t}\E\lim_{t\to0}\frac1h\bigg[\max\{u(\theta,x-y)
      +\tilde V(s,y,\bar t+h)\}\\
    &\qquad{}-\max\{u(\theta,x-y)
      +\tilde V(s,y,\bar t)\}\bigg]\lambda e^{\lambda(t-s)}\:ds\\
    &\qquad{}+\lim_{h\to0}\frac1h
      \int_{\bar t}^{\bar t+h}\E\max\{u(\theta,x-y)
      +\tilde V(s,y,\bar t+h)\}\lambda e^{\lambda(t-s)}\:ds\\
    &\qquad{}-\lim_{h\to0}\frac1h
      \int_{\bar t}^{\bar t+h}\E\max\{u(\theta,x+\bar x-y)
      +\tilde V(s,y,\bar t)\}\lambda e^{\lambda(t-s)}\:ds\\[\medskipamount]
  &=\int_t^{\bar t}\E\bigg[\frac d{d\bar t}\max\{u(\theta,x-y)
      +\tilde V(s,y,\bar t)\}\bigg]
      \lambda e^{\lambda(t-s)}\:ds\\[\medskipamount]
    &\qquad{}+\E\max_{0\leq y\leq x}\{u(\theta,x-y)+V(\bar t,\bar x+y)\}
      \lambda e^{\lambda(t-\bar t)}\\
    &\qquad{}-\E\max_{0\leq y\leq x+\bar x}
      \{u(\theta,x+\bar x-y)+V(\bar t,y)\}\lambda e^{\lambda(t-\bar t)}.
\end{align*}
Applying Lemma~\ref{discrete-envelope} gives us
\begin{align*}
\frac d{d\bar t}\tilde V(t,x+\bar x\chi_{[\bar t,T]}(t))
  &=\int_t^{\bar t}\E\left[\frac{d\tilde V}{d\bar t}
      \,\bigg|_{s,y^*}\right]
      \lambda e^{\lambda(t-s)}\:ds\\
    &\qquad{}+\E\max_{0\leq y\leq x}\{u(\theta,x-y)+V(\bar t,\bar x+y)\}
      \lambda e^{\lambda(t-\bar t)}\\
    &\qquad{}-\E\max_{0\leq y\leq x+\bar x}
      \{u(\theta,x+\bar x-y)+V(\bar t,y)\}\lambda e^{\lambda(t-\bar t)}.
\end{align*}
Differentiating both sides with respect to $t$ gives us \eq{differential-eq-tilde-V}. For reference, that equation is
\[
\frac{d}{dt}\left(\frac{d\tilde V}{d\bar t}\,\bigg|_{t,x}\right)=
  \lambda\frac{d\tilde V}{d\bar t}\,\bigg|_{t,x}-
  \lambda\E\left[\frac{d\tilde V}{d\bar t}\,\bigg|_{t,y^*}\right]
  e^{\lambda(t-\bar t)}
\]
with initial condition
\begin{align*}
\frac{d\tilde V}{d\bar t}\,\bigg|_{\bar t,x}&=
    \lambda\E\max_{0\leq y\leq x}\{u(\theta,x-y)+V(\bar t,\bar x+y)\}
  -\lambda\E\max_{0\leq y\leq x+\bar x}
    \{u(\theta,x+\bar x-y)+V(\bar t,y)\}
\end{align*}
Consider first the case where $x=0$. Then $y^*=0$, and
\[
\frac{d}{dt}\left(\frac{d\tilde V}{d\bar t}\,\bigg|_{t,0}\right)=
    \lambda\frac{d\tilde V}{d\bar t}\,\bigg|_{t,0}-
    \lambda\frac{d\tilde V}{d\bar t}\,\bigg|_{t,0}
    e^{\lambda(t-\bar t)}
  =\lambda(1-e^{\lambda(t-\bar t)})
    \frac{d\tilde V}{d\bar t}\,\bigg|_{t,0}.
\]
In this case, $d\tilde V/d\bar t=0$ is a nullcline of the differential equation, so $d\tilde V/d\bar t$ never changes sign. The initial condition is
\[
\frac{d\tilde V}{d\bar t}\,\bigg|_{\bar t,0}=
    \lambda V(\bar t,\bar x)
  -\lambda\E\max_{0\leq y\leq \bar x}
    \{u(\theta,\bar x-y)+V(\bar t,y)\}=V_t(\bar t,\bar x)<0,
\]
so $d\tilde V/d\bar t$ is always negative when $x=0$.

Now we induct on $x$. Assume $d\tilde V/d\bar t$ is always negative when the agent starts with $x-1$ or fewer resources. Consider the situation where $d\tilde V/d\bar t=0$. Then
\[
\frac{d}{dt}\left(\frac{d\tilde V}{d\bar t}\,\bigg|_{t,x}\right)=
  -\lambda\E\left[\frac{d\tilde V}{d\bar t}\,\bigg|_{t,y^*}\right]
  e^{\lambda(t-\bar t)},
\]
which is positive by the induction hypothesis since $y^*<x$ with positive probability. It follows that the positive real numbers is an attracting region for $d\tilde V/d\bar t$. On the other hand, suppose $y^*$ maximizes $u(\theta,x-y)+V(\bar t,\bar x+y)$. Then the initial condition becomes
\begin{align*}
\frac{d\tilde V}{d\bar t}\,\bigg|_{\bar t,x}&=
      \lambda\E[u(\theta,x-y^*)+V(\bar t,\bar x+y^*)]\\
    &\qquad{}-\lambda\E\max_{0\leq y\leq x+\bar x}
      \{u(\theta,x+\bar x-y)+V(\bar t,y)\}\\
  &\leq\lambda\E[u(\theta,x-y^*)+V(\bar t,\bar x+y^*)]\\
    &\qquad{}-\lambda\E[u(\theta,x+\bar x-[y^*+\bar x])+
      V(\bar t,y^*+\bar x)]=0.
\end{align*}
It follows that when the agent has $x$ resources and when $t<\bar t$, we have $d\tilde V/d\bar t<0$. So the result holds for all $x$.
\end{proof}

\th{log-concave-property}

\begin{proof}
By definition, a function $f$ is log-concave if $\log f$ is concave. We have
\[
\frac{d^2}{dx^2}\log f=\frac d{dx}\left(\frac{f'}f\right)
  =\frac{f''f-(f')^2}{f^2}<0.
\]
This means $f''f<(f')^2$, and composing both sides with $f^{-1}$ on the right gives us
\begin{align*}
xf''\circ f^{-1}(x)&<[f'\circ f^{-1}(x)]^2\\
\frac d{dx}f'\circ f^{-1}(x)=\frac{f''\circ f^{-1}(x)}{f'\circ f^{-1}(x)}
  &<\frac{f'\circ f^{-1}(x)}x.
\end{align*}
It follows that at any point $x$, the derivative of $f'\circ f^{-1}$ is smaller than the slope of the segment from the origin to $(x,f'\circ f^{-1}(x))$. I claim this means that the graph of $f'\circ f^{-1}$ lies above every segment from the origin to a point on the graph of $f'\circ f^{-1}$. Suppose we have points $x$ and $y$ such that $0<y<x$, and $f'\circ f^{-1}(y)$ is below the segment between the origin and $f'\circ f^{-1}(x)$. Define
\[
z=\sup\left\{p\mathpunct{\colon} 0<p<x\text{ and }f'\circ f^{-1}(p)<
  \frac{p}{x}f'\circ f^{-1}(x)\right\}.
\]
This set is nonempty (because of $y$) and bounded above by $x$, so $z$ exists. It must be the case that 
\[
f'\circ f^{-1}(z)=\frac{z}{x}f'\circ f^{-1}(x).
\]
If $z=x$, this equation is true trivially. For $z<x$, if $f'\circ f^{-1}(z)<(z/x)f'\circ f^{-1}(x)$, then this inequality holds in a neighborhood of $z$, so $z$ cannot be the supremum. It follows that we have equality at $z$. At the same time, there exists a sequence $(p_i)$ that converges to $z$ such that this inequality holds at each $p_i$. This means that
\begin{align*}
\frac d{dz}f'\circ f^{-1}(z)
  &=\lim_{i\to\infty}\frac{f'\circ f^{-1}(z)-f'\circ f^{-1}(p_i)}{z-p_i}\\
  &\geq\lim_{i\to\infty}
    \frac{f'\circ f^{-1}(z)-(p_i/x)f'\circ f^{-1}(x)}{z-p_i}\\
  &=\lim_{i\to\infty}
    \frac{(z/x)f'\circ f^{-1}(x)-(p_i/x)f'\circ f^{-1}(x)}{z-p_i}\\
  &=\frac{f'\circ f^{-1}(x)}x=\frac{f'\circ f^{-1}(z)}z.
\end{align*}
Log-concavity implies the opposite inequality is true, so by contrapositive, $y$ cannot exist. We conclude that the graph of $f'\circ f^{-1}$ does in fact lie above every segment between the origin and any point on the graph of $f'\circ f^{-1}$. In mathematical symbols, we write that property as $af'\circ f^{-1}(x)<f'\circ f^{-1}(ax)$, which is exactly what we want to prove.
\end{proof}

\th{cutoffs-approaching}

\begin{proof}
We know that
\begin{align*}
\zeta(\phi_{i,j})[\mu(i-j+1)-\mu(i-j)]
  &=\zeta(\phi_{i+1,j})[\mu(i-j+2)-\mu(i-j+1)]\\
\zeta(\phi_{i,j})&=\zeta(\phi_{i+1,j})
  \frac{\mu(i-j+2)-\mu(i-j+1)}{\mu(i-j+1)-\mu(i-j)}\\
\phi_{i,j}&=\zeta^{-1}\left(\zeta(\phi_{i+1,j})
  \frac{\mu(i-j+2)-\mu(i-j+1)}{\mu(i-j+1)-\mu(i-j)}\right).
\end{align*}
If we write $a=\Delta\mu(i-j+1)/\Delta\mu(i-j)$, then $0<a<1$, and we have $\phi_{i,j}=\zeta^{-1}(a\zeta(\phi_{i+1,j}))$. Now consider $\phi_{i+1,j}-\phi_{i,j}$. Differentiating gives us
\begin{align*}
\frac d{dt}\phi_{i+1,j}-\phi_{i,j}
  &=\frac d{dt}\phi_{i+1,j}-\zeta^{-1}(a\zeta(\phi_{i+1,j}))\\
  &=\phi_{i+1,j}'-
    \frac{a\zeta'(\phi_{i+1,j})}
      {\zeta'\circ\zeta^{-1}(a\zeta(\phi_{i+1,j}))}
    \phi_{i+1,j}'\\
  &=\phi_{i+1,j}'\left[1-\frac{a\zeta'(\phi_{i+1,j})}
      {\zeta'\circ\zeta^{-1}(a\zeta(\phi_{i+1,j}))}\right].
\end{align*}
From Lemma~\ref{log-concave-property}, the term inside brackets is positive if $\zeta$ is log-concave. Because $\phi_{i+1,j}$ is decreasing, its derivative is negative, so the product must be negative. It follows that $\phi_{i+1,j}-\phi_{i,j}$ is strictly decreasing.
\end{proof}

\th{more-concave-w-marginal}

\begin{proof}
Theorem~\ref{phi-psi-compare-single} establishes this result for the case $i=0$. We establish the general case by induction. Suppose the result is true up to $i-1$. By performing a variable substitution on each integral, we can rewrite \eq{decomposition} as
\begin{align}
V_{i+1}'-V_i'&=-\sum_{k=0}^{i-1}
  \int_{\min\{V_{i-k+1}-V_{i-k},\mu(k+1)-\mu(k)\}}
    ^{\min\{V_{i-k}-V_{i-k-1},\mu(k+1)-\mu(k)\}}
    \left[1-\zeta^{-1}
      \left(\frac{y}{\mu(k+1)-\mu(k)}\right)\right]\:dy\nonumber\\
  &\qquad{}-\int_{\min\{V_1,\mu(i+1)-\mu(i)\}}
    ^{\mu(i+1)-\mu(i)}
    \left[1-\zeta^{-1}\left(\frac{y}{\mu(i+1)-\mu(i)}\right)\right]\:dy
\label{decomposition-change-var}
\end{align}
Suppose $k_1<k_2$. Then
\begin{align*}
\Delta\mu(k_1)&>\Delta\mu(k_2)\\
\frac y{\Delta\mu(k_1)}&<\frac y{\Delta\mu(k_2)}\\
1-\zeta^{-1}\left(\frac y{\Delta\mu(k_1)}\right)&>
  1-\zeta^{-1}\left(\frac y{\Delta\mu(k_2)}\right).
\end{align*}
This means that in each successive addend from \eq{decomposition-change-var}, the integrand gets smaller. At the same time, in each successive addend, the bounds $\mu(k+1)-\mu(k)$ in the limits of integration get smaller. It follows that for $k>0$, increasing $V_{i-k}-V_{i-k-1}$ makes the right side of \eq{decomposition-change-var} stay the same or become more negative. At the same time, replacing $\zeta$ with a more concave function $\xi$ also makes the right side of \eq{decomposition-change-var} more negative because
\begin{align*}
\zeta&<\xi\\
\zeta^{-1}&>\xi^{-1}\\
1-\zeta^{-1}&<1-\xi^{-1}.
\end{align*}
Now consider \eq{decomposition-change-var} to be a differential equation for $V_{i+1}'-V_i'$. We also have an analogous differential equation for $W_{i+1}'-W_i'$, and by the inductive hypothesis, when $k>0$, $V_{i-k+1}-V_{i-k}<W_{i-k+1}-W_{i-k}$. Suppose at time $t$, $V_{i+1}-V_{i}=W_{i+1}-W_{i}$. Then
\begin{align*}
V_{i+1}'-V_i'&=-\sum_{k=0}^{i-1}
  \int_{\min\{V_{i-k+1}-V_{i-k},\mu(k+1)-\mu(k)\}}
    ^{\min\{V_{i-k}-V_{i-k-1},\mu(k+1)-\mu(k)\}}
    \left[1-\zeta^{-1}
      \left(\frac{y}{\mu(k+1)-\mu(k)}\right)\right]\:dy\\
  &\qquad{}-\int_{\min\{V_1,\mu(i+1)-\mu(i)\}}
    ^{\mu(i+1)-\mu(i)}
    \left[1-\zeta^{-1}\left(\frac{y}{\mu(i+1)-\mu(i)}\right)\right]\:dy\\
&\geq-\sum_{k=0}^{i-1}
  \int_{\min\{W_{i-k+1}-W_{i-k},\mu(k+1)-\mu(k)\}}
    ^{\min\{W_{i-k}-W_{i-k-1},\mu(k+1)-\mu(k)\}}
    \left[1-\zeta^{-1}
      \left(\frac{y}{\mu(k+1)-\mu(k)}\right)\right]\:dy\nonumber\\
  &\qquad{}-\int_{\min\{W_1,\mu(i+1)-\mu(i)\}}
    ^{\mu(i+1)-\mu(i)}
    \left[1-\zeta^{-1}\left(\frac{y}{\mu(i+1)-\mu(i)}\right)\right]\:dy\\
&>-\sum_{k=0}^{i-1}
  \int_{\min\{W_{i-k+1}-W_{i-k},\mu(k+1)-\mu(k)\}}
    ^{\min\{W_{i-k}-W_{i-k-1},\mu(k+1)-\mu(k)\}}
    \left[1-\xi^{-1}
      \left(\frac{y}{\mu(k+1)-\mu(k)}\right)\right]\:dy\nonumber\\
  &\qquad{}-\int_{\min\{W_1,\mu(i+1)-\mu(i)\}}
    ^{\mu(i+1)-\mu(i)}
    \left[1-\xi^{-1}\left(\frac{y}{\mu(i+1)-\mu(i)}\right)\right]\:dy
    =W_{i+1}'-W_i'.
\end{align*}
The result follows from Lemma~\ref{differential-eq-compare}.
\end{proof}

\end{document}